\documentclass[fleqn,preprint]{elsarticle}
\AtBeginDocument{\setlength{\mathindent}{1pt}}

\usepackage{booktabs}
\usepackage{longtable}
\usepackage[flushleft]{threeparttable}
\usepackage[flushleft]{threeparttablex}
\usepackage{multirow}
\usepackage{multicol}
\usepackage{amsmath,amssymb,amsfonts}
\usepackage{algorithm}
\usepackage{array}
\usepackage{textcomp}
\usepackage{stfloats}
\usepackage{verbatim}

\usepackage{eurosym}
\usepackage{acronym}
\usepackage[mathscr]{eucal}
\usepackage{xcolor}
\usepackage{enumitem}
\usepackage{accents}
\usepackage{natbib}
\usepackage{geometry}
\usepackage{graphicx}
\usepackage{txfonts}
\usepackage{hyperref}
\usepackage{float}
\usepackage{tcolorbox}
\tcbuselibrary{skins,breakable}

\biboptions{sort&compress}

\usepackage{subcaption} % For subfigure support
\usepackage{caption}    % For better control over captions
\usepackage{url}
\usepackage{breakurl}

\usepackage{makecell}   % Allows text wrapping in table cells

\usepackage{pifont} % Include in the preamble

\usepackage{rotating}
\usepackage[nameinlink]{cleveref}

\usepackage{tabularx}

\usepackage{tikz}
\usetikzlibrary{shapes.geometric, arrows.meta, positioning}
\tikzstyle{block} = [rectangle, draw, text centered, minimum height=1.2em, font=\small, rounded corners, fill=blue!10]
\tikzstyle{decision} = [diamond, draw, text centered, inner sep=0pt, minimum width=2cm, font=\small]
\tikzstyle{input} = [rectangle, draw, text centered, font=\small, fill=gray!20]
\tikzstyle{uncertain} = [ellipse, draw, dashed, fill=red!10, font=\small]
\tikzstyle{arrow} = [thick,->,>=stealth]

\tikzset{
  panel label/.style={
    anchor=north west,
    font=\small,
    fill=white, fill opacity=0.75, text opacity=1,
    inner sep=2pt,
    xshift=3mm, yshift=-3mm
  }
}

\usepackage{algpseudocode}

\usetikzlibrary{arrows.meta,positioning,shapes.geometric,fit,calc}

\newcommand{\ubar}[1]{\underaccent{\bar}{#1}}

\usepackage{nomencl}

\makenomenclature

\usepackage{xstring}
\usepackage{etoolbox}
\renewcommand\nomgroup[1]{%
  \item[\bfseries
  \ifstrequal{#1}{I}{Indexes and Sets}{%
  \ifstrequal{#1}{P}{Parameters}{%
  \ifstrequal{#1}{V}{\textcolor{black}{Variables}}{%
  \ifstrequal{#1}{W}{Binary Variables}{%
  \ifstrequal{#1}{Z}{Vectors}{}}}}}%
]}

\graphicspath{{figs/}} 

\acrodef{sb}[SB]{Single-Bound}
\acrodef{mb}[MB]{Multi-Bound}
\acrodef{plmbro}[PL-MBRO]{Probability-Layer MBRO}
\acrodef{mbro}[MBRO]{Multi-Bound Robust Optimization}
\acrodef{dam}[DAM]{Day-Ahead Market}
\acrodef{srm}[SRM]{Secondary Reserve Market}
\acrodef{ess}[ESS]{Energy Storage System}
\acrodef{res}[RES]{Renewable Energy Sources}
\acrodef{der}[DER]{Distributed Energy Resources}
\acrodef{ndrs}[ND-RES]{Non-Dispatchable RES}
\acrodef{drs}[D-RES]{Dispatchable RES}
\acrodef{vpp}[VPP]{Virtual Power Plant}
\acrodef{pv}[PV]{Photovoltaic}
\acrodef{milp}[MILP]{Mixed Integer Linear Programming}
\acrodef{rVPP}[RVPP]{Renewable-Based VPP}
\acrodef{rvpp}[RVPP]{Renewable-Based Virtual Power Plant}
\acrodef{gams}[GAMS]{General Algebraic Modeling System}
\acrodef{so}[SO]{Stochastic Optimization}
\acrodef{ro}[RO]{Robust Optimization}
\acrodef{aro}[ARO]{Adaptive Robust Optimization}
\acrodef{saro}[SARO]{Stochastic Adaptive RO}
\acrodef{dro}[DRO]{Distributionally Robust Optimization}
\acrodef{ccg}[C\&CG]{Column \& Constraint Generation}
\acrodef{ev}[EV]{Electric Vehicle}
\acrodef{igdt}[IGDT]{Information Gap Decision Theory}
\acrodef{cvar}[CVaR]{Conditional Value-at-Risk}
\acrodef{vab}[VaB]{Value-at-Best}
\acrodef{soc}[SoC]{State of Charge}
\acrodef{pvt}[PVT]{Photovoltaic-Thermal}
\acrodef{woa}[WOA]{Whale Optimization Algorithm}
\acrodef{dr}[DR]{Demand Response}
\acrodef{chp}[CHP]{Combined Heat and Power}
\acrodef{cco}[CCO]{Chance-Constrained Optimization}
\acrodef{rec}[REC]{Renewable Energy Certificate}
\acrodef{cer}[CER]{Carbon Emission Right}
\acrodef{mc}[MC]{Marginal Contributions}
\acrodef{dp}[DP]{Dynamic Programming}

\journal{International Journal of Electrical Power \& Energy Systems}

\ExplSyntaxOn
\NewDocumentCommand{\instringTF}{mmmm}
{\oleks_instring:nnnn { #1 } { #2 } { #3 } { #4 }}
\tl_new:N \l__oleks_instring_test_tl
\cs_new_protected:Nn \oleks_instring:nnnn
{
    \tl_set:Nn \l__oleks_instring_test_tl { #1 }
    \regex_match:nnTF { \u{l__oleks_instring_test_tl} } { #2 } { #3 } { #4 }
}
\ExplSyntaxOff
\let\mybibitem\bibitem
\renewcommand{\bibitem}[1]{%
    \instringTF{Refblack}{#1}
    {\color{black}\mybibitem{#1}}
    {\color{black}\mybibitem{#1}}%
}

\begin{document}

\begin{frontmatter}

\title{Probability-Layer Clustering for Bound-and-Budget Calibration in Multi-Bound Robust Bidding of Renewable-Based Virtual Power Plants}

\author{Hadi Nemati*}
\author{{\'A}lvaro Ortega}
\author{Enrique Lobato}
\author{Luis Rouco}

\address{Comillas Pontifical University, Institute for Research in Technology, Madrid, Spain}    

\address{*Corresponding author}
\address {E-mail address: hnemati@comillas.edu}

\begin{abstract}

\acp{rvpp} that aggregate dispatchable and non-dispatchable renewable energy sources, storage, and flexible demand that bid in the day-ahead and secondary reserve markets must cope with uncertainties in market prices, renewable generation, and demand. \ac{mbro} is well suited to this setting because it hedges adverse realizations through several deviation layers while keeping the final bidding problem as a tractable, scenario-free mixed-integer linear program. Its performance, however, depends on how the deviation bounds and uncertainty budgets are calibrated. Existing \ac{mbro} implementations commonly rely on fixed percentile-based bounds and externally specified budget vectors, usually examined through sensitivity analysis. This paper proposes a probability-layer \ac{mbro} approach, referred to as \acs{plmbro}, that constructs parameter-specific deviation bounds from historical data conditional on selected risk-attitude quantiles and automatically derives the corresponding uncertainty budgets from empirical cluster probabilities. For each uncertain parameter, $k$-medians clustering partitions the historical daily profiles into representative probability layers; the cluster median defines the balanced representative, while favorable and adverse within-cluster percentiles define the optimistic and pessimistic representatives. The empirical cluster probabilities are mapped to integer full-horizon uncertainty budgets over the 96 quarter-hourly periods, while the robust inner problem determines the periods in which the corresponding deviation layers are most adverse to the schedule. Compared with a percentile-based \ac{mbro} benchmark, in-sample and out-of-sample tests on an \ac{rvpp} bidding in both markets show that \acs{plmbro} consistently achieves better realized economic outcomes across the optimistic, balanced, and pessimistic strategies, with improvements of up to 92.7\% in-sample and 57.3\% out-of-sample, measured relative to the absolute value of the corresponding benchmark realized profit.

\end{abstract}

\end{frontmatter}

\section*{Keywords}
Renewable-based virtual power plant, Probability-layer multi-bound robust optimization, $k$-medians clustering, Data-driven uncertainty-set calibration, Day-ahead market, Secondary reserve market

\vspace{2cm}

\begin{tcolorbox}[breakable,colback=white,colframe=black,
title=Nomenclature,width=\textwidth]

\begin{footnotesize}

% =========================================================
\noindent \textbf{Sets and Indices}
\vspace{.5em}

\renewcommand{\arraystretch}{1.15}
\begin{tabularx}{\textwidth}{@{}lX@{}}

$t \in \mathscr{T}$ \hspace{3.5mm}
& Scheduling periods over the daily optimization horizon \\

$r \in \mathscr{R}$
& Non-dispatchable renewable generation units \\

$c \in \mathscr{C}$
& Dispatchable renewable generation units \\

$s \in \mathscr{S}$
& Energy storage systems \\

$d \in \mathscr{D}$
& Flexible electricity demands \\

$u \in \mathscr{U}$
& Unit-side uncertain quantities, comprising renewable availability and demand \\

$w \in \mathscr{W}$
& Reserve-providing assets, comprising non-dispatchable and dispatchable renewable units, storage systems, and flexible demands \\

$k \in \mathscr{K}$
& Uncertainty layer (bound) of the multi-bound robust uncertainty set \\

$\Xi^{O}$
& Set of outer-level bidding, scheduling, and operational decision variables \\

$\Xi^{I}$
& Set of variables introduced by the inner-problem reformulation \\

\end{tabularx}

\vspace{.5em}

% =========================================================
\noindent \textbf{Parameters}
\vspace{.5em}

\begin{tabularx}{\textwidth}{@{}lX@{}}

$C_c,C_r,C_s$ \hspace{3mm}
& Marginal operating-cost coefficients of dispatchable, non-dispatchable, and storage resources
\hfill [€/MWh] \\

$P_{r,t},P_{d,t}$
& Available non-dispatchable generation and demand reference in period $t$
\hfill [MW] \\

$\bar P,\ubar P$
& Upper and lower power limits; the corresponding asset is identified by the subscript
\hfill [MW] \\

$\bar E,\ubar E$
& Upper and lower energy limits; the corresponding asset is identified by the subscript
\hfill [MWh] \\

$R_w^{\uparrow},R_w^{\downarrow}$
& Upward and downward ramp-rate capabilities of reserve-providing asset $w$
\hfill [MW/min] \\

$T^{SR}$
& Secondary-reserve activation time
\hfill [min] \\

$\beta_w^{\uparrow},\beta_w^{\downarrow}$
& Maximum fractions of rated power available for upward and downward reserve
\hfill [\%] \\

$\Gamma_k$
& Full-horizon uncertainty budget associated with layer $k$
\hfill [--] \\

$\Delta t$
& Duration of one scheduling period
\hfill [h] \\

$\eta_s^{+},\eta_s^{-}$
& Charging and discharging efficiencies of storage system $s$
\hfill [\%] \\

$\lambda_t^{DA}$
& Day-ahead energy price in period $t$
\hfill [€/MWh] \\

$\lambda_t^{SR,\uparrow},\lambda_t^{SR,\downarrow}$
& Upward and downward secondary-reserve capacity prices in period $t$
\hfill [€/MW] \\

$M$
& Valid upper-bound constant used in the Big-$M$ linearization
\hfill [MW] \\

\end{tabularx}

\vspace{.5em}

% =========================================================
\noindent \textbf{Decision Variables}
\vspace{.5em}

\begin{tabularx}{\textwidth}{@{}lX@{}}

$e_{s,t}$ \hspace{6mm}
& Stored energy of storage system $s$ at the end of period $t$
\hfill [MWh] \\

$p_t^{DA}$
& Net day-ahead market position of the RVPP in period $t$; positive values denote sales and negative values purchases
\hfill [MW] \\

$r_t^{SR,\uparrow},r_t^{SR,\downarrow}$
& Upward and downward secondary-reserve capacities offered by the RVPP
\hfill [MW] \\

$\sigma_s^{\uparrow},\sigma_s^{\downarrow}$
& Fractions of the usable storage-energy range allocated to upward and downward reserve
\hfill [\%] \\

$v_{c,t}$
& Binary commitment status of dispatchable unit $c$ during period $t$
\hfill [--] \\

$v_{s,t}$
& Binary storage operating-mode variable; $1$ denotes charging and $0$ denotes discharging
\hfill [--] \\

\end{tabularx}

\vspace{.5em}

% =========================================================
\noindent \textbf{Superscripts and Accents}
\vspace{.5em}

\begin{tabularx}{\textwidth}{@{}lX@{}}

$\vartheta^{DA}$ \hspace{4.7mm}
& Quantity associated with the day-ahead market \\

$\vartheta^{SR}$
& Quantity associated with the secondary reserve market \\

$\vartheta^{M}$
& Market-side quantity or vector associated with day-ahead and secondary-reserve products \\

$\vartheta^\uparrow,\vartheta^\downarrow$
& Upward and downward reserve directions, respectively \\

$\vartheta^+,\vartheta^-$
& Charging-side and discharging-side quantities, respectively \\

$\tilde{\vartheta}$
& Nominal or reference value of an uncertain quantity \\

$\vartheta^{*}$
& Optimal value of a variable \\

$\hat{\vartheta},\check{\vartheta}$
& Upper-side and lower-side deviation magnitudes relative to the nominal reference, respectively \\

$\bar{\vartheta},\ubar{\vartheta}$
& Upper and lower admissible limits, respectively \\

\end{tabularx}

\vspace{-1em}
\end{footnotesize}
\end{tcolorbox}

\acresetall %resets all acronym definitions
\section{Introduction}

\subsection{Motivation}

The increasing share of variable \ac{res} is making the coordinated use of distributed flexibility increasingly important in electricity markets. Rather than operating heterogeneous resources independently, a \ac{rvpp} can aggregate \acp{ndrs}, controllable \acp{drs}, \acp{ess}, and flexible demand and coordinate their complementary capabilities for energy trading and reserve provision. Recent studies show that \acp{rvpp} are increasingly being designed for multi-market participation while simultaneously facing uncertainty in renewable production, electricity demand, and market prices~\cite{zare2026systematic,xu2026bidding,kong2026coordinated}. This challenge becomes more pronounced as electricity markets move toward finer temporal resolution. In Spain, for example, the transition from 24 hourly intervals to 96 quarter-hourly intervals increases the temporal detail at which energy positions, resource availability, demand, and market prices must be represented~\cite{EC_Energy_Market,REE2025,omie_homepage}. Consequently, effective \ac{rvpp} participation requires uncertainty-management approaches that remain economically meaningful when many interdependent market and operational decisions must be made over a high-resolution daily horizon.

From the \ac{rvpp} operator's perspective, however, the difficulty is not simply the larger number of scheduling periods. \ac{dam} energy bids and upward and downward \ac{srm} reserve offers must be determined before the corresponding market prices, renewable availability, and demand are realized. Deviations from the anticipated conditions can affect both the profitability and the feasibility of the resulting market position: renewable shortfalls or unexpectedly high demand may create energy imbalances, while changing a unit's scheduled energy alters the operating headroom available for upward or downward reserve. At quarter-hourly resolution, the uncertain parameters may also exhibit different patterns of severity and occurrence throughout the day. Moderate deviations may arise relatively frequently, whereas extreme renewable shortfalls, demand excursions, or price movements may occur much less often; moreover, these patterns can differ substantially among parameters. Applying the same deviation structure or horizon-wide protection pattern across such distinct conditions can therefore produce a level of conservatism that is poorly aligned with the historical data. This motivates an uncertainty representation that distinguishes different deviation levels and reflects how frequently their underlying operating regimes have been observed.

These considerations make the calibration of the uncertainty representation an important part of the bidding problem. Budgeted \ac{ro} is attractive because it can protect market decisions without introducing a large scenario set into the final optimization model, but its classical single-bound form associates the protected periods with one deviation magnitude~\cite{bertsimas04}. \ac{mbro} extends this structure by assigning separate budgets to several deviation layers, allowing moderate and more severe deviations to be represented differently~\cite{busing2012new,nemati2026multi}. Nevertheless, the operational meaning of those layers depends on how their magnitudes and full-horizon budgets are obtained. In the existing percentile-based \ac{mbro} formulation for quarter-hourly \ac{rvpp} bidding, the bounds are formed from predefined percentile levels and the budgets are selected externally and examined through sensitivity analysis~\cite{nemati2026multi}. Such an analysis reveals how alternative settings affect the solution, but it does not itself infer a parameter-specific allocation of the scheduling horizon among the representative deviation levels. This creates a need for a reproducible calibration that relates the multi-bound structure to observed daily profiles while preserving tractability and keeping empirical occurrence separate from the operator's risk attitude.

\vspace{2cm}

\subsection{Literature Review}

Recent \ac{vpp} research has expanded from isolated energy scheduling toward coordinated participation in multiple market products. Recent work documents the increasing interaction of \acp{vpp} with energy, reserve, regulation, flexibility, capacity, and carbon markets, together with the coordination of increasingly heterogeneous resource portfolios~\cite{zare2026systematic}. Yang et al.~\cite{yang2023optimal} study day-ahead bidding for a renewable-based portfolio, whereas Shang et al.~\cite{shang2025uncertainty} coordinate participation in day-ahead energy, frequency-modulation, and peaking markets and subsequently allocate the resulting profit among the participating resources. Du et al.~\cite{du2025optimal} construct physically and economically feasible operating regions before embedding the aggregated \ac{vpp} in a bi-level joint energy-and-reserve bidding and clearing model. Pleite et al.~\cite{pleite2026bidding} develop day-ahead bidding and offering curves for a \ac{vpp} participating in a manual frequency-restoration-reserve-based demand-response program. Chang and Yang~\cite{chang2025game} instead emphasize risk-aware wholesale-market bidding together with internal electricity trading among heterogeneous resource aggregators. Although these studies broaden market participation and resource coordination, uncertainty is either neglected or limited to selected sources. Among the reviewed studies, none jointly models market-price, renewable-generation, and demand uncertainties within a unified energy-and-reserve bidding framework.

As the scope of \ac{vpp} participation expands, \acp{vpp} must increasingly account for uncertainty in renewable generation, demand, and energy and reserve market prices. Scenario-based \ac{so} provides a common way to represent these uncertain quantities through a set of possible realizations. Falabretti et al.~\cite{falabretti2023scheduling} develop an integrated stochastic scheduling model for a renewable-based \ac{vpp} with e-mobility, while Mei et al.~\cite{mei2024two} use a two-step spot-market model in which energy-storage scheduling responds to renewable-generation forecast errors. Nokandi et al.~\cite{nokandi2023three} extend the decision structure to three stages for joint energy-and-reserve scheduling while incorporating a local intraday demand-response exchange. Risk considerations are incorporated more explicitly by Chang and Yang~\cite{chang2025game}, who use classified real-time electricity-price scenarios and \ac{cvar} in day-ahead wholesale-market bidding, and by Pleite et al.~\cite{pleite2026bidding}, who represent day-ahead prices and reserve-program activation through scenarios within a risk-averse stochastic formulation. Gela et al.~\cite{gela2026metamodel} jointly represent wind, load, and electricity-price uncertainty through retained scenarios in a stochastic Stackelberg coordination model. More recent multi-market formulations broaden the uncertain quantities further. Bian et al.~\cite{bian2026cvar} use scenarios and \ac{cvar} to account for renewable-output and market-price risk in coordinated energy--regulation--carbon bidding, whereas Li et al.~\cite{li2026risk} consider renewable and market-price uncertainty together with uncertain schedulable boundaries and compliance willingness of controllable loads in capacity, energy, and ancillary-service markets, using $k$-medoids clustering to construct representative scenarios. Kong et al.~\cite{kong2026coordinated} combine forecasting of wind and solar generation, demand, and electricity prices with scenario generation and approximate \ac{dp} for coordinated day-ahead and real-time energy-and-ancillary-service decisions. These studies show that stochastic formulations can flexibly represent several uncertainty sources and market decisions; however, their performance depends on the quality of the selected scenarios and their assigned probabilities, while a larger scenario set generally increases computational requirements.

The composition and coordinated use of flexible resources constitute another important dimension of \ac{vpp} scheduling. Energy storage is increasingly combined with renewable generation and flexible demand to provide temporal flexibility and support participation in different market products. Feng et al.~\cite{feng2024optimal} integrate electric vehicles and \acp{ess} with renewable generation, while Wang et al.~\cite{wang2024optimal} coordinate aggregated user-side distributed storage and photovoltaic resources. Wang et al.~\cite{wang2024two} likewise combine photovoltaic generation, energy storage, and charging-station demand within a price-maker \ac{vpp}. Flexibility has also been extended beyond energy trading: Afzali et al.~\cite{afzali2025flexibility} coordinate local energy communities in energy and flexibility markets, whereas Esfahani et al.~\cite{esfahani2024stochastic} aggregate backup batteries for participation in the frequency-regulation market. Siqin et al.~\cite{siqin2025two} coordinate a multi-energy \ac{vpp} across energy, reserve, and carbon markets, and Xu et al.~\cite{xu2026bidding} exploit wind--solar complementarity together with flexible resource scheduling in coordinated energy and frequency markets. At the operational level, Wang et al.~\cite{wang2026drmpc} further illustrate the importance of temporal flexibility by modeling workload shifting and the dynamic evolution of electric-vehicle charging within a \ac{vpp}-enabled integrated energy system. These studies show that the economic value of a \ac{vpp} depends not only on the markets it accesses but also on how the temporal and cross-market flexibility of its heterogeneous resources is represented. As scheduling intervals become shorter, these intertemporal and energy--reserve couplings must be represented with increasing temporal detail.

Another closely related issue is how available information is converted into the parameters required by uncertainty-aware scheduling models. Different studies use data for different purposes. Wu et al.~\cite{wu2024two} use \ac{dro} to construct an ambiguity-set representation for correlated wind and solar uncertainty, while Ma et al.~\cite{ma2025data} develop data-driven uncertainty intervals for \ac{vpp} bidding under price, renewable-generation, and demand uncertainty. Historical or simulated data have also been used to learn operational policies: Lv et al.~\cite{lv2026drl} train a deep-reinforcement-learning policy for scheduling under renewable variability, whereas Bo et al.~\cite{bo2026adaptive} use reinforcement learning to adjust the look-ahead horizon and robustness parameters according to the operating state. Other formulations rely on model-specific calibration parameters: \ac{igdt} uses risk-attitude settings to define uncertainty envelopes~\cite{zhu2025low,shafiekhani2022optimal}, whereas \ac{cco} uses prescribed probability or confidence levels~\cite{feng2025optimal,du2025optimal}. Robust formulations may instead use forecast deviations, externally selected budgets, or sensitivity analysis~\cite{liu2025two,li2023robust,nemati2025segan}. Thus, historical information is increasingly incorporated into \ac{vpp} scheduling, but the way in which it is translated into optimization inputs differs substantially across applications. For an \ac{mbro} model, the specific calibration problem is to determine not only representative deviation magnitudes but also the horizon-wide allocation assigned to the different deviation levels.

Temporal granularity has become another important consideration as electricity-market operation moves from hourly toward sub-hourly intervals. Existing \ac{vpp} studies incorporate this finer resolution in different parts of the decision process. Li et al.~\cite{li2025internal} develop a high-resolution internal--external coordinated bidding strategy for day-ahead spot and peaking ancillary-service markets, while Yuanyuan et al.~\cite{yuanyuan2023distributionally} consider fine-grained market declarations together with flexible-ramping products. Wang et al.~\cite{wang2023optimal} combine an hourly day-ahead energy schedule with a finer reserve-market representation, illustrating that energy and ancillary-service decisions need not necessarily share the same temporal resolution. Mixed-resolution structures also appear in more recent studies: Bian et al.~\cite{bian2026cvar} use hourly day-ahead decisions followed by sub-hourly real-time adjustments, Chang and Yang~\cite{chang2025game} employ finer internal resource schedules while retaining an hourly wholesale-market horizon, and Kong et al.~\cite{kong2026coordinated} use high-resolution historical information within coordinated day-ahead and real-time decision making. These studies demonstrate the growing importance of finer temporal representation, although the increased resolution is often introduced only for selected operational or market stages. In the present setting, energy and secondary-reserve bids, resource schedules, and the associated uncertain parameters are represented consistently throughout the complete sub-hourly daily horizon.

Within set-based approaches, the multi-band uncertainty construction of B\"using and D'Andreagiovanni proposed in~\cite{busing2012new} is particularly relevant because it replaces a single adverse deviation level with several bounded layers. Its recent adaptation to high-resolution \ac{rvpp} energy-and-reserve bidding showed that multiple deviation bounds can improve the robustness--profit trade-off while retaining a tractable single-level \ac{milp}~\cite{nemati2026multi}. In that formulation, however, the deviation magnitudes are obtained from predefined percentile pairs and the corresponding budget vector is supplied externally and examined through sensitivity analysis. The percentile-based \ac{mbro} formulation is therefore the closest benchmark for the present study because the market model, resource portfolio, and underlying multi-bound robust structure can be retained while changing only the calibration procedure. The remaining methodological question is how historical daily profiles can be used to determine both the representative deviation layers and their full-horizon budgets in a reproducible and parameter-specific manner.

%Against this background, the proposed \ac{plmbro} approach introduces a data-driven calibration layer for the existing multi-bound robust bidding model. Historical daily profiles are organized into parameter-specific probability layers using $k$-medians; within-cluster representatives determine the deviation magnitudes, while empirical cluster probabilities determine the corresponding full-horizon budgets. The risk-attitude-dependent representatives modify the severity of the bounds without changing these probability-based budgets. Consequently, the proposed contribution concerns the empirical construction and calibration of the multi-bound uncertainty set while preserving the underlying single-level \ac{milp} bidding formulation. Table~\ref{table:Literature} summarizes how the reviewed studies relate to the market scope, temporal resolution, uncertainty sources, and uncertainty-model construction considered in this work.

Against this background, the proposed \ac{plmbro} approach introduces a data-driven calibration layer for the existing multi-bound robust bidding model. Historical daily profiles are grouped into parameter-specific clusters using $k$-medians, providing two complementary sources of information for uncertainty-set calibration. The empirical frequency of each cluster is used to determine its corresponding full-horizon uncertainty budget, while representative values within each cluster are used to construct the deviation bounds. Specifically, the cluster median represents the balanced strategy, whereas favorable and adverse within-cluster quantiles are selected for the optimistic and pessimistic strategies, respectively. In this way, the operator's risk attitude affects the severity of the deviation bounds without changing the probability-based uncertainty budgets. Consequently, the proposed contribution concerns the empirical construction and calibration of the multi-bound uncertainty set while preserving the underlying single-level \ac{milp} bidding formulation. Table~\ref{table:Literature} summarizes how the reviewed studies relate to the market scope, temporal resolution, uncertainty sources, and uncertainty-model construction considered in this work.

\begin{table*}[t!]
  \centering
  \caption{Representative comparison of uncertainty-aware \acs{vpp} scheduling and market studies.}
  \scriptsize
  \setlength{\tabcolsep}{1.5pt}
  \renewcommand{\arraystretch}{1.8}
  \vspace{-1em}
  \begin{threeparttable}
  \begin{tabular}{@{}p{1.2cm}cccp{2.2cm}p{3.5cm}p{3.5cm}c@{}}
    \toprule
    \textbf{Ref.}
    & \makecell{\textbf{Energy \&}\\\textbf{reserve}}
    & \makecell{\textbf{Sub-hourly}\\\textbf{resolution}}
    & \makecell{\textbf{Multi-source}\\\textbf{uncertainty}}
    & \makecell{\textbf{Uncertainty}\\\textbf{representation}}
    & \makecell{\textbf{Uncertainty-input}\\\textbf{basis}}
    & \makecell{\textbf{Conservatism/risk}\\\textbf{control}}
    & \textbf{Method} \\
    \midrule

    \cite{liu2025two}
    &
    &
    &
    & Single-bound set
    & Forecast-deviation bounds
    & External budget/sensitivity
    & \acs{ro} \\[0.15em]

    \cite{wu2024two,feng2024optimal}
    &
    &
    &
    & Ambiguity set
    & Historical samples
    & Ambiguity radius
    & \acs{dro} \\[0.15em]

    \cite{falabretti2023scheduling}
    &
    &
    &
    & Scenarios
    & Monte Carlo sampling
    & Scenario probabilities
    & \acs{so} \\[0.15em]

    \cite{mei2024two}
    &
    &
    &
    & Scenarios
    & Forecast-error distributions
    & Scenario probabilities
    & \acs{so} \\[0.15em]

    \cite{chang2025game}
    &
    & $\checkmark$
    &
    & Scenarios
    & \makecell[l]{Historical data+SVM scenario\\ reduction}
    & \acs{cvar} parameters
    & \acs{so} \\[0.15em]

    \cite{gela2026metamodel}
    &
    &
    & $\checkmark$
    & Scenarios
    & \makecell[l]{Historical statistics+LHS\\+$k$-means reduction}
    & Scenario probabilities
    & \acs{so} \\[0.15em]

    \cite{wang2024optimal,wang2024two}
    &
    &
    & $\checkmark$
    & Ambiguity set
    & Historical samples
    & Ambiguity/risk parameters
    & \acs{dro} \\[0.15em]

    \cite{zhu2025low,shafiekhani2022optimal}
    &
    &
    & $\checkmark$
    & \makecell[l]{Information-gap \\envelope}
    & Forecast-deviation envelopes
    & Risk-attitude parameter
    & \acs{igdt} \\[0.15em]

    \cite{afzali2025flexibility}
    &
    &
    & $\checkmark$
    & Scenarios/bounds
    & \makecell[l]{Scenario-based data\\+worst-case availability}
    & \makecell[l]{Scenario probabilities/\\robust setting}
    & \acs{so}--\acs{ro} \\[0.15em]

    \cite{ma2025data}
    &
    &
    & $\checkmark$
    & Single-bound set
    & Data-driven interval construction
    & External robustness setting
    & \acs{ro} \\[0.15em]

    \cite{yang2023optimal,shang2025uncertainty}
    & $\checkmark$
    &
    &
    & Probabilistic/interval
    & \makecell[l]{Forecast-error distributions/\\probability intervals}
    & Probability/confidence parameters
    & -- \\[0.15em]

    \cite{feng2025optimal,du2025optimal}
    & $\checkmark$
    &
    &
    & Chance constraints
    & \makecell[l]{Probability distributions/\\intervals}
    & Confidence level
    & \acs{cco} \\[0.15em]

    \cite{siqin2025two}
    & $\checkmark$
    &
    & $\checkmark$
    & Ambiguity set
    & Historical samples
    & Ambiguity/risk settings
    & \acs{dro} \\[0.15em]

    \cite{xu2026bidding}
    & $\checkmark$
    &
    & $\checkmark$
    & Ambiguity-based/multi-model
    & \makecell[l]{Multi-feature forecasts\\+multi-stage scenario generation}
    & Ambiguity/risk settings
    & \acs{dro} \\[0.15em]

    \cite{esfahani2024stochastic}
    & $\checkmark$
    &
    &
    & Scenarios/bounds
    & Monte Carlo+backward reduction
    & Probabilities/robust settings
    & \acs{so}--\acs{ro} \\[0.15em]

    \cite{nokandi2023three}
    & $\checkmark$
    &
    &
    & Scenarios
    & Probabilistic scenario tree
    & Scenario probabilities
    & \acs{so} \\[0.15em]

    \cite{pleite2026bidding}
    & $\checkmark$
    &
    &
    & Scenarios
    & AR(1)+normal-error sampling
    & \acs{cvar} parameters
    & \acs{so} \\[0.15em]

    \cite{li2026risk}
    & $\checkmark$
    &
    & $\checkmark$
    & Scenarios
    & Monte Carlo+$k$-medoids clustering
    & \acs{cvar} confidence/risk weight
    & \acs{so} \\[0.15em]

    \cite{bian2026cvar}
    & $\checkmark$
    & $\checkmark$
    & $\checkmark$
    & Scenarios
    & \makecell[l]{LSTM+ARMA error model\\+Monte Carlo+$k$-medoids}
    & Data-driven risk coefficient
    & \acs{so} \\[0.15em]

    \cite{nemati2025segan}
    & $\checkmark$
    &
    & $\checkmark$
    & Single-bound set
    & Percentile-based bounds
    & External budget/sensitivity
    & \acs{ro} \\[0.15em]

    \cite{kong2026coordinated}
    & $\checkmark$
    &
    & $\checkmark$
    & Scenarios
    & \makecell[l]{Forecast-based LHS\\+Wasserstein backward reduction}
    & Operator risk preference
    & DP \\[0.15em]

    \cite{li2025internal}
    & $\checkmark$
    & $\checkmark$
    &
    & Ambiguity set
    & Historical samples
    & Ambiguity radius
    & \acs{dro} \\[0.15em]

    \cite{yuanyuan2023distributionally}
    & $\checkmark$
    & $\checkmark$
    &
    & Ambiguity set
    & Historical samples
    & Ambiguity radius
    & \acs{dro} \\[0.15em]

    \cite{wang2023optimal}
    & $\checkmark$
    & $\checkmark$
    &
    & Single-bound set
    & Forecast-deviation bounds
    & External budget/sensitivity
    & \acs{ro} \\[0.15em]

    \cite{nemati2026multi}
    & $\checkmark$
    & $\checkmark$
    & $\checkmark$
    & Multi-bound set
    & Predefined percentile bounds
    & External budgets/sensitivity
    & \acs{mbro} \\[0.15em]

    \textbf{This paper}
    & $\checkmark$
    & $\checkmark$
    & $\checkmark$
    & Multi-bound set
    & \makecell[l]{Historical profiles+$k$-medians \\clustering}
    & \makecell[l]{Cluster-probability budgets/\\quantile levels}
    & \acs{plmbro} \\

    \bottomrule
  \end{tabular}

    \begin{tablenotes}
  \footnotesize
  \item A check mark in \textbf{Energy \& reserve} indicates joint participation in energy and reserve/ancillary-service markets. A check mark in \textbf{Sub-hourly resolution} indicates that at least one bidding, scheduling, or operational decision layer uses a time resolution shorter than one hour; sub-hourly input data alone are not sufficient. A check mark in \textbf{Multi-source uncertainty} indicates that both market-price uncertainty and unit-side uncertainty (renewable generation and/or demand) are considered. The \textbf{Uncertainty-input basis} column summarizes how the uncertainty representation is constructed or populated rather than the specific uncertain quantities considered. Grouped references share the indicated broad characteristics. External budgets denote robustness parameters selected outside the data-driven calibration procedure. SVM denotes support vector machine, LHS denotes Latin hypercube sampling, and LSTM denotes long short-term memory.
    \end{tablenotes}

  \end{threeparttable}
  \label{table:Literature}
\end{table*}

\subsection{Paper Contributions}

The main contributions of this paper are as follows:

\begin{itemize}

\item \textbf{Empirical construction of parameter-specific probability layers:} Historical daily trajectories of the \ac{dam} and \ac{srm} prices, \ac{ndrs} production, and demand are clustered separately using $k$-medians. The complete-sample median provides the nominal reference, cluster medians define the balanced representatives, and direction-dependent within-cluster quantiles generate the optimistic and pessimistic deviation magnitudes. This construction allows each uncertain parameter to retain its own empirical profile structure instead of being forced into common global percentile bands.

\item \textbf{Probability-based calibration of full-horizon uncertainty budgets:} The empirical probability of each cluster is mapped to the 96-period scheduling horizon. The resulting integer budgets are parameter-specific, sum to the complete horizon, and remain fixed across the three uncertainty-handling strategies. Risk attitude is therefore represented through the severity of the within-cluster representatives rather than by manually retuning the budgets for every strategy.

\item \textbf{Integration with a tractable joint energy-and-reserve bidding model:} The calibrated deviations and budgets are embedded in the existing multi-bound robust structure for simultaneous \ac{dam} and \ac{srm} participation. The final problem remains a single-level \ac{milp}, so the proposed calibration improves the empirical construction of the uncertainty set without replacing the robust market model or introducing a scenario set into the optimization.

%\item \textbf{Evaluation of bidding-stage and realized economic performance:} 

\end{itemize}

The optimized bids are evaluated under realized historical conditions both within the calibration sample and on temporally held-out days, across optimistic, balanced, and pessimistic strategies and using bidding-stage and settlement-based economic metrics. This assessment evaluates whether the economic performance of the bidding decisions obtained with the proposed calibration persists under unseen historical realizations.

The remainder of the paper is organized as follows: Section~\ref{sec:Problem_description} specifies the \ac{rvpp} portfolio, market interface, and decision sequence studied in this work. Section~\ref{sec:Formulation} develops the deterministic and the proposed probability-layer multi-bound bidding formulations and then defines the data-driven calibration procedure. Section~\ref{sec:Case_Studies} presents the calibrated uncertainty sets together with the bidding-stage, in-sample, and held-out-data results. Finally, conclusions and future extensions of this work are summarized in Section~\ref{sec:Conclusion}.

\newpage

\section{Problem Description}
\label{sec:Problem_description}

\begin{figure}[t!]
\centering
{\includegraphics[width=\linewidth]{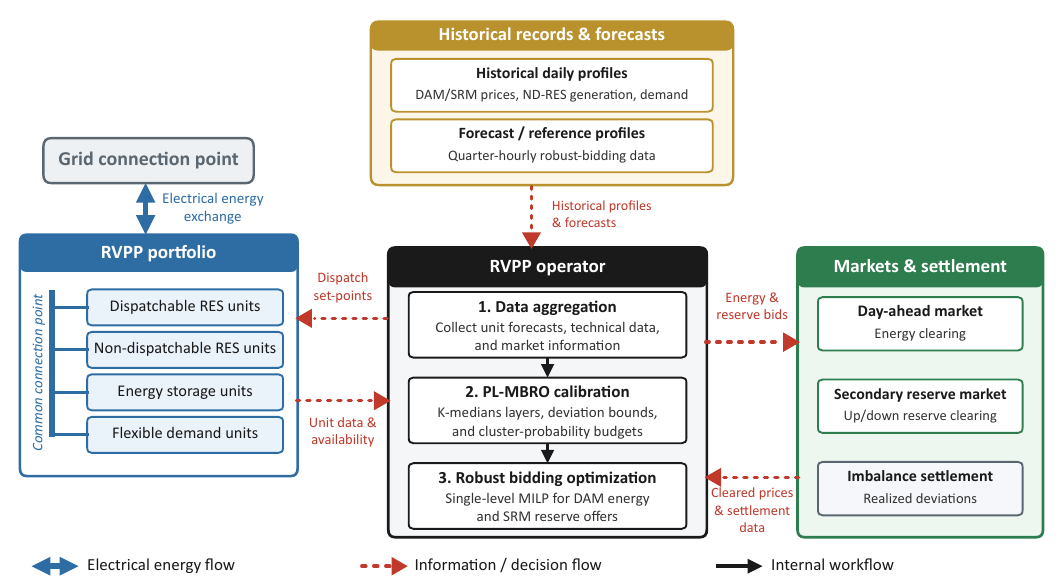}}
\caption{Operational framework of the proposed \ac{rvpp} and its data-driven \ac{plmbro} calibration process.}
\label{fig:RVPP_Layout}
\end{figure}

Figure~\ref{fig:RVPP_Layout} illustrates the operational framework considered in this work. The \ac{rvpp} acts as a single market participant that aggregates a heterogeneous portfolio of flexible and uncertain resources, including \acp{drs}, \acp{ndrs}, \acp{ess}, and/or flexible demands. The portfolio exchanges electrical energy with the power system through one or more grid connection points, while the \ac{rvpp} operator coordinates the internal resources and represents the aggregated portfolio in the electricity markets. The operator receives technical data, availability information, and historical records of the uncertain parameters, including \ac{dam} and \ac{srm} prices, \ac{ndrs} generation, and demand. These data are used to construct the probability-layer multi-bound uncertainty set, whose deviation bounds and full-horizon budgets are calibrated as detailed in~\Cref{subsec:Calibration_MB}. Using the resulting uncertainty set, the operator solves the robust bidding problem and submits energy bids to the \ac{dam} together with upward and downward reserve offers to the \ac{srm}. The \ac{rvpp} is modeled as a price-taker; accordingly, offer prices are not treated as strategic decision variables, and zero-price offers are used to promote market acceptance. This assumption is reasonable given the relatively small size of the \ac{rvpp} considered relative to the power system and the low marginal cost of its renewable resources. The optimization jointly determines the market positions and the corresponding unit-level schedules, which are implemented following market clearing, while deviations between committed and realized operation are subsequently accounted for through market settlement.

The main challenge is that the \ac{rvpp} must commit its \ac{dam} and \ac{srm} positions before market prices, renewable availability, and demand are realized. These uncertain quantities affect the revenues obtained from energy and reserve products and may also generate imbalance-settlement costs when the realized portfolio position deviates from the committed schedule. In the proposed framework, uncertainty management is organized into two steps. 

First, historical daily profiles are processed using $k$-medians clustering to identify representative probability layers for each uncertain parameter. Strategy-dependent within-cluster representatives determine the corresponding deviation bounds, while the empirical cluster probabilities determine their full-horizon uncertainty budgets. Second, the calibrated bounds and budgets are embedded in the robust bidding model, whose inner problem determines the periods in which the available deviation layers are most adverse to the \ac{rvpp} schedule. The operator's risk attitude is represented through the choice of the within-cluster representatives: the balanced strategy uses the cluster median, whereas the optimistic and pessimistic strategies use favorable and adverse within-cluster percentiles, respectively. 

The proposed \ac{plmbro} approach thus preserves the tractable multi-bound robust-bidding structure while replacing predefined percentile-based bounds and externally specified uncertainty budgets with a data-driven, parameter-specific calibration procedure aligned with quarter-hourly market operation.

\section{Formulation}
\label{sec:Formulation}
This section presents the optimization model for the \ac{rvpp}. \Cref{subsec:Deterministic_Formulation} presents the deterministic bidding problem for participation in the \ac{dam} and \ac{srm}, following standard formulations~\cite{ma2025data, zhu2025low, wang2023optimal, nemati2025segan}. \Cref{subsec:Multi-Bound_Robust} then develops the corresponding \ac{plmbro} model, which protects the bids against uncertainty in market prices, \ac{ndrs} generation, and demand~\cite{nemati2026multi}. The data-driven calibration of the uncertainty bounds and budgets, which constitutes the central contribution of this paper, is presented in~\Cref{subsec:Calibration_MB}.

\subsection{Deterministic Problem}
\label{subsec:Deterministic_Formulation}

\subsubsection{Objective Function}
\label{subsubsec:Objective_Function}

The objective function in~\eqref{RVPP: Obj_Deterministic} maximizes the daily operating profit of the \ac{rvpp}. It comprises revenues from \ac{dam} energy trading and upward and downward reserve provision in the \ac{srm}, minus the operating costs of the dispatchable units, non-dispatchable units, and storage systems.

\begingroup
\allowdisplaybreaks
\vspace{-.5em}
\begin{align} \label{RVPP: Obj_Deterministic}
&\mathop {\max}\limits_{\mathbf{x}}
\sum\limits_{t \in \mathscr{T}}
{{\lambda_t^{DA}p_t^{DA}\Delta t}}
+
\sum\limits_{t \in \mathscr{T}}
{\left[
{{\lambda_t^{{SR,\uparrow}}r_t^{SR,\uparrow}}
+
{\lambda_t^{{SR,\downarrow}}r_t^{SR,\downarrow}}}
\right]}
-\sum\limits_{t \in \mathscr{T}}
{\sum\limits_{c \in \mathscr{C}}
{{C_c}p_{c,t}\Delta t}}
-\sum\limits_{t \in \mathscr{T}}
{\sum\limits_{r \in \mathscr{R}}
{C_rp_{r,t}\Delta t}}
-\sum\limits_{t \in \mathscr{T}}
{\sum\limits_{s \in \mathscr{S}}
{C_sp_{s,t}^{-}\Delta t}}
\end{align}
\vspace{.1em}
\endgroup

\subsubsection{Power Balance Constraint}
\label{subsubsec:Supply_Demand_Constraints}

The portfolio must remain balanced in every time period under each possible secondary-reserve activation state. These states are represented compactly by the reserve-state vectors $\boldsymbol{r}_{t}^{SR}=\{r_{t}^{SR,\uparrow},-r_{t}^{SR,\downarrow},0\}$ for the \ac{rvpp} and $\boldsymbol{r}_{w,t}=\{r_{w,t}^{\uparrow},-r_{w,t}^{\downarrow},0\}$ for each asset $w\in\mathscr{W}$ (\acp{ndrs}, \acp{ess}, \acp{drs}, and demands). Their three components correspond, respectively, to upward activation, downward activation, and no activation. Consequently, the vector equality~\eqref{cons:Supply-Demand1} simultaneously enforces the three corresponding power-balance conditions~\cite{nemati2025segan}.

\begingroup
\allowdisplaybreaks
\vspace{-.8em}
%\begin{subequations}
\begin{align}
    &\!\sum\limits_{r \in \mathscr{R}} \left[ p_{r,t} \!+ \boldsymbol{r}_{r,t} \right] \!+ \!\sum\limits_{s \in \mathscr{S}} \left[ p_{s,t} \!+ \boldsymbol{r}_{s,t} \right] \!+ \!\sum\limits_{c \in \mathscr{C}} \left[ p_{c,t} \!+ \boldsymbol{r}_{c,t} \right] %+\sum\limits_{s \in \mathscr{S}} \left[ p_{s,t} + \boldsymbol{r}_{s,t} \right]
     - \sum\limits_{d \in \mathscr{D}} \left[ p_{d,t} - \boldsymbol{r}_{d,t} \right] = p_{t}^{DA}+ \boldsymbol{r}_{t}^{SR}~;
    & 
    \forall t \label{cons:Supply-Demand1}
\end{align}
%\label{RVPP:Supply-Demand}
%\end{subequations}
\vspace{.1em}
\endgroup

%\vspace{1em}
\subsubsection{Unit-level Operational Constraints}
\label{subsubsec:DRES_Constraints}

The operational constraints for each technology are summarized below. Each block also specifies how the corresponding unit contributes energy and reserve.

\noindent\textbf{Dispatchable units.} \ac{drs}, such as hydroelectric power plants, can be dispatched within their operating limits and can provide both energy and reserve. Constraints~\eqref{Deterministic: DRES1}--\eqref{Deterministic: DRES2} bound the joint energy--reserve schedule through the commitment variable $v_{c,t}$, whereas~\eqref{Deterministic: DRES3} limits daily energy production, for example, under seasonal hydro-management requirements. Minimum up- and down-time constraints are omitted for brevity and can be formulated as in~\cite{carrion2006computationally}.

\begingroup
\allowdisplaybreaks
\begin{subequations}
\vspace{-.9em}
\begin{align}
    &p_{c,t} + r_{c,t}^{\uparrow} \leq \bar P_{c} v_{c,t}~; 
    &
    \forall c, t \label{Deterministic: DRES1} \\
    & \ubar P_{c} v_{c,t} \leq p_{c,t} - r_{c,t}^{\downarrow}~; 
    &
    \forall c, t \label{Deterministic: DRES2} \\
    &\sum\limits_{t \in \mathscr{T}} \left[{p_{c,t}\Delta t} + r_{c,t}^{\uparrow}\Delta t \right] \le \bar E_{c}~;
    &
    \forall c \label{Deterministic: DRES3} 
    \end{align}    
\label{Deterministic: DRES}
\end{subequations}
\vspace{-.8em}
\endgroup

\noindent\textbf{Non-dispatchable units.} \ac{ndrs}, such as wind and solar \ac{pv} units, have weather-dependent upper bounds on output but can still provide reserve by adjusting curtailment. Downward reserve is delivered by reducing output, whereas upward reserve is available when a unit is pre-curtailed below its available generation potential~\cite{zhang2022frequency, yin2021state}. Constraint~\eqref{cons: NDRES1} limits scheduled production plus upward reserve to the forecast available power $P_{r,t}$, whereas~\eqref{cons: NDRES2} imposes the lower feasibility bound after downward-reserve activation.
\begingroup
\allowdisplaybreaks
\begin{subequations}
%\vspace{-.8em}
\begin{align}
    & p_{r,t}+r_{r,t}^{\uparrow} \leq P_{r,t}~; 
    & 
    \forall r,t \label{cons: NDRES1} \\ 
    & \ubar P_{r} \le p_{r,t}-r_{r,t}^{\downarrow}~; 
    & 
    \forall r,t  \label{cons: NDRES2} 
    \end{align}
\label{RVPP: NDRES}
\end{subequations}
\vspace{-1em}
\endgroup

\noindent\textbf{Flexible demand.} Flexible demand can provide reserve while its consumption remains subject to uncertainty~\cite{feng2025optimal}. Constraints~\eqref{cons: Demand1}--\eqref{cons: Demand2} define the feasible joint consumption--reserve schedule, where $P_{d,t}$ denotes the demand forecast. Constraint~\eqref{cons: Demand3} enforces the minimum energy requirement over the scheduling horizon.
\begingroup
%\vspace{-.5em}
\allowdisplaybreaks
\begin{subequations}
\begin{align}
    & P_{d,t} \le p_{d,t} - r_{d,t}^{\uparrow}~;
    &
    \forall d,t  \label{cons: Demand1} \\
    & p_{d,t} + r_{d,t}^{\downarrow} \le \bar P_{d}~;
    & 
    \forall d,t  \label{cons: Demand2} \\
    & \ubar E_{d} \le \sum\limits_{t \in \mathscr{T}} \left[{p_{d,t}\Delta t} - r_{d,t}^{\uparrow}\Delta t \right]~;
    &
    \forall d \label{cons: Demand3}
\end{align}
\label{RVPP: Demand}
\end{subequations}
\vspace{-.5em}
\endgroup

\noindent\textbf{Energy storage.} Each \ac{ess} can perform energy arbitrage and provide reserve, as modeled by~\eqref{Deterministic: ES}~\cite{nemati2025segan}. Constraints~\eqref{Deterministic: ES2}--\eqref{Deterministic: ES5} bound the charging and discharging schedules and their associated reserve capacities, with the binary operating-state variable $v_{s,t}$ preventing simultaneous charging and discharging. Equations~\eqref{Deterministic: ES6}--\eqref{Deterministic: ES8} define net power and aggregate upward and downward reserve. Equation~\eqref{Deterministic: ES9} updates the state of charge, whereas~\eqref{Deterministic: ES10} imposes a cyclic boundary condition by equating the state of charge in the initial and final periods; \eqref{Deterministic: ES11}--\eqref{Deterministic: ES12} set aside storage headroom for reserve via $\sigma_{s}^{\uparrow}$ and $\sigma_{s}^{\downarrow}$; and~\eqref{Deterministic: ES13} shrinks the feasible energy range accordingly.
\begingroup
\allowdisplaybreaks
\begin{subequations}
\begin{align}
    &\ubar {P}_{s}^{+} v_{s,t} \leq p_{s,t}^{+} - r_{s,t}^{+,\uparrow}~; 
    &\forall s, t \label{Deterministic: ES2} \\
    & p_{s,t}^{+} + r_{s,t}^{+,\downarrow} \leq \bar {P}_{s}^{+} v_{s,t}~;
    &\forall s, t \label{Deterministic: ES3} \\
    &p_{s,t}^{-} + r_{s,t}^{-,\uparrow} \leq \bar {P}_{s}^{-} \left( 1 - v_{s,t} \right)~; 
    &\forall s, t \label{Deterministic: ES4} \\
    &\ubar {P}_{s}^{-} \left( 1 - v_{s,t} \right) \leq p_{s,t}^{-} - r_{s,t}^{-,\downarrow}~; 
    &\forall s, t \label{Deterministic: ES5} \\
    &p_{s,t} = p_{s,t}^{-} - p_{s,t}^{+}~; 
    &\forall s, t \label{Deterministic: ES6} \\
    &r_{s,t}^{\uparrow} = r_{s,t}^{+,\uparrow} + r_{s,t}^{-,\uparrow}~; 
    &\forall s, t \label{Deterministic: ES7} \\
    &r_{s,t}^{\downarrow} = r_{s,t}^{+,\downarrow} + r_{s,t}^{-,\downarrow}~; 
    &\forall s, t \label{Deterministic: ES8} \\
    & e_{s,t} = e_{s,t-1} + p_{s,t}^{+} \eta_s^{+} \Delta t - \frac{p_{s,t}^{-} \Delta t}{\eta_{s}^{-}}~;
    &\forall s, t \label{Deterministic: ES9} \\
    & e_{s,0} = e_{s, |\mathscr{T}|}~;% = \alpha_s, 
    &\forall s \label{Deterministic: ES10}\\
    & \sum_{t \in \mathscr{T}} \frac{r_{s,t}^{\uparrow} \Delta t} {\eta_s^{-}} \leq \sigma_{s}^{\uparrow} \left( \bar E_s - \ubar E_s \right)~; 
    &\forall s \label{Deterministic: ES11} \\
    & \sum_{t \in \mathscr{T}} r_{s,t}^{\downarrow} \eta_s^{+} \Delta t \leq \sigma_{s}^{\downarrow} \left( \bar E_s - \ubar E_s \right)~; 
    &\forall s \label{Deterministic: ES12} \\
    &\ubar E_s + \sigma_s^{\uparrow} \left( \bar E_s - \ubar E_s \right) \leq e_{s,t} \leq \bar E_{s} - \sigma_{s}^{\downarrow} \left( \bar E_{s} - \ubar E_{s} \right);& 
    \forall s, t \label{Deterministic: ES13}   
    %
    %& v_{s,t} \in \{0,1\}; 
    %&\forall s, t \label{Deterministic: ES14}
\end{align}
\label{Deterministic: ES}
\end{subequations}
%\vspace{-.6em}
\endgroup

\noindent\textbf{Reserve provision.} Constraints~\eqref{Deterministic: Reserve} impose common reserve limits across all portfolio assets. Equations~\eqref{Deterministic: Reserve1}--\eqref{Deterministic: Reserve2} limit upward and downward reserve according to each asset's ramp rate and the secondary-reserve activation time. Equations~\eqref{Deterministic: Reserve3}--\eqref{Deterministic: Reserve4} impose additional limits based on the unit rating and the operator-specified reserve-provision factors~\cite{nemati2025segan}.
\begingroup
\allowdisplaybreaks
%\vspace{-1em}
\begin{subequations}
\begin{align}
    & r_{w,t}^{\uparrow} \le T^{SR} R_{w}^{\uparrow}~;
    & 
    \forall w,t  \label{Deterministic: Reserve1} \\
    & r_{w,t}^{\downarrow} \le T^{SR} R_{w}^{\downarrow}~;
    & 
    \forall w,t  \label{Deterministic: Reserve2} \\
    & r_{w,t}^{\uparrow} \le \beta^{\uparrow}_w \bar P_{w}~;
    & 
    \forall w,t  \label{Deterministic: Reserve3} \\
    & r_{w,t}^{\downarrow} \le \beta^{\downarrow}_w \bar P_{w}~;
    & 
    \forall w,t  \label{Deterministic: Reserve4}
    \end{align}    
\label{Deterministic: Reserve}
\vspace{-2em}
\end{subequations}
\endgroup
\subsection{Probability-Layer Multi-Bound Robust Model}
\label{subsec:Multi-Bound_Robust}

The deterministic formulation in~\Cref{subsec:Deterministic_Formulation} assumes that market prices, \ac{ndrs} availability, and demand are known with certainty. The \ac{plmbro} model relaxes this assumption by protecting the \ac{rvpp} against adverse realizations of \ac{dam} and \ac{srm} prices, renewable generation, and demand. As in conventional robust optimization, the model focuses on adverse deviations that reduce profitability or restrict feasibility, while favorable realizations are accounted for when evaluating the resulting bids under actual historical conditions. Each uncertain parameter is represented by a set of deviation bounds indexed by $k\in\mathscr{K}$. Each bound $k$ is associated with a representative data cluster, and its corresponding uncertainty budget determines how many periods may attain that deviation level over the scheduling horizon. In the proposed \ac{plmbro} approach, the deviation bounds and their budgets are calibrated using the cluster representatives and their empirical probabilities, as described in~\Cref{subsec:Calibration_MB}. Consequently, the level of protection is distributed among the deviation bounds according to their observed frequencies rather than being determined through externally selected uncertainty budgets.

The model adopts a bi-level robust structure. The outer problem determines the bidding and scheduling decisions, whereas the inner problems identify the most adverse market-price, renewable-generation, and demand realizations permitted by the corresponding multi-bound uncertainty sets. The model is first presented in bi-level form, after which the inner linear problems are reformulated using duality to obtain a single-level \ac{milp}. To avoid repeating equivalent expressions for each uncertain market price and asset parameter, related quantities are stacked into vectors and combined using element-wise operations. The notation and bi-level formulation are introduced in~\Cref{subsubsec:Bilevel_formulation}, the inner protection problems are presented in~\Cref{subsubsec:Inner_Reformulation}, and the resulting tractable formulation is provided in~\Cref{subsubsec:MILP_Formulation}.

\subsubsection{Bi-Level Formulation}
\label{subsubsec:Bilevel_formulation}

We first define the uncertainty affecting the objective function. The three market products---\ac{dam} energy and upward/downward \ac{srm} reserve---are stacked so that their price contributions can be treated jointly. For each period $t$ and bound $k$, the nominal price vector and the magnitudes of the negative and positive deviations are
\begin{equation*}
\tilde{\boldsymbol{\lambda}}_t^{M}\!\triangleq\!
\begin{bmatrix}\tilde{\lambda}_t^{DA}\\[0.2mm] \tilde{\lambda}_t^{SR,\uparrow}\\[0.2mm] \tilde{\lambda}_t^{SR,\downarrow}\end{bmatrix}\!,\;\;
\check{\boldsymbol{\lambda}}_{k,t}^{M}\!\triangleq\!
\begin{bmatrix}\check{\lambda}_{k,t}^{DA}\\[0.2mm] \check{\lambda}_{k,t}^{SR,\uparrow}\\[0.2mm] \check{\lambda}_{k,t}^{SR,\downarrow}\end{bmatrix}\!,\;\;
\hat{\boldsymbol{\lambda}}_{k,t}^{M}\!\triangleq\!
\begin{bmatrix}\hat{\lambda}_{k,t}^{DA}\\[0.2mm] 0\\[0.2mm] 0\end{bmatrix}\!,\;\;
\boldsymbol{\rho}_{k,t}^{M}\!\triangleq\!
\begin{bmatrix}\check{\lambda}_{k,t}^{DA}/\hat{\lambda}_{k,t}^{DA}\\[0.2mm] 0\\[0.2mm] 0\end{bmatrix}\!.
\end{equation*}

The \ac{dam} price admits deviations in both directions because the adverse realization depends on the net market position: a price decrease is adverse when the \ac{rvpp} sells energy, whereas a price increase is adverse when it buys energy. Accordingly, $\lambda_t^{DA}\!\in\![\tilde{\lambda}_t^{DA}\!-\!\check{\lambda}_{k,t}^{DA},\,\tilde{\lambda}_t^{DA}\!+\!\hat{\lambda}_{k,t}^{DA}]$, where the two deviation magnitudes may differ. Because the \ac{rvpp} only sells reserve, only negative reserve-price deviations are adverse, and the corresponding positive-deviation entries are set to zero. The vector $\boldsymbol{\rho}_{k,t}^{M}$ captures the asymmetry between the negative and positive \ac{dam} price deviations, while its reserve components are zero because reserve-price uncertainty is one-sided. The scheduled market quantities, the vector used to isolate the energy position, and the vector used to represent the asymmetric energy-price effect are
\begin{equation*}
\boldsymbol{q}_t^{M}\!\triangleq\!
\begin{bmatrix}p_t^{DA}\Delta t\\[0.2mm] r_t^{SR,\uparrow}\\[0.2mm] r_t^{SR,\downarrow}\end{bmatrix}\!,\qquad
\boldsymbol{e}_t^{M}\!\triangleq\!
\begin{bmatrix}p_t^{DA}\Delta t\\[0.2mm] 0\\[0.2mm] 0\end{bmatrix}\!,\qquad
\boldsymbol{y}_{k,t}^{M}\!\triangleq\!
\begin{bmatrix}y_{k,t}^{DA}\\[0.2mm] r_t^{SR,\uparrow}\\[0.2mm] r_t^{SR,\downarrow}\end{bmatrix}\!,
\end{equation*}
The vector $\boldsymbol{q}_t^{M}$ contains the scheduled market quantities evaluated at the nominal prices. The vector $\boldsymbol{e}_t^{M}$ isolates the \ac{dam} energy component to enable the separate treatment of its positive and negative price deviations, while its reserve entries are set to zero because reserve-price deviations are unidirectional. The vector $\boldsymbol{y}_{k,t}^{M}$ contains the quantities exposed to the adverse price deviations associated with bound $k$. In particular, $y_{k,t}^{DA}$ is a nonnegative auxiliary variable that represents the magnitude of the energy position exposed to the adverse \ac{dam} price deviation. By accounting separately for the positive deviation $\hat{\lambda}_{k,t}^{DA}$ and the negative deviation $\check{\lambda}_{k,t}^{DA}$, this variable identifies the adverse price realization according to whether the \ac{rvpp} buys or sells energy in period $t$.

Using the notation above, the outer-level objective is written as~\eqref{Uncertainty: Obj} over the decision set $\Xi^{O}$. The outer-level variables are collected in $\Xi^{O}\!\triangleq\!\{\mathbf{x},y_{k,t}^{DA}\}_{k\in\mathscr{K},\,t\in\mathscr{T}}$, where $\mathbf{x}$ contains the bidding, scheduling, and operational variables of the \ac{rvpp}. The first term represents the market revenue evaluated at nominal prices. The outer-level decisions are gathered in $\mathbf{x}$. The term $\boldsymbol{c}\,\mathbf{x}$ represents the deterministic unit operation cost in the objective function. The vector $\boldsymbol{\Theta}^{M}$, obtained from the inner price-protection problem, represents the worst-case revenue loss caused by adverse market-price deviations; therefore, $\mathbf{1}^{\top}\boldsymbol{\Theta}^{M}$ is the total robust cost deducted from the nominal profit. Constraint~\eqref{Uncertainty: Asymetric_DAprice} links $y_{k,t}^{DA}$ to the sign of the energy position, so that the model applies a negative price deviation to sold energy and a positive price deviation to purchased energy. The symbol $\odot$ denotes element-wise (Hadamard) multiplication.

\begingroup
\allowdisplaybreaks
\begin{subequations}
\begin{align}
\mathop{\max}\limits_{\Xi^{O}}\;
&\sum\limits_{t \in \mathscr{T}} \big(\boldsymbol{q}_t^{M}\big)^{\!\top}\tilde{\boldsymbol{\lambda}}_t^{M}
\;-\; \boldsymbol{c}\,\mathbf{x}
\;-\; \mathbf{1}^{\top}\boldsymbol{\Theta}^{M}~; \label{Uncertainty: Obj}\\
    &\text{s.t.}\;\;-\boldsymbol{\rho}_{k,t}^{M}\odot \boldsymbol{y}_{k,t}^{M} \le \boldsymbol{e}_{t}^{M} \le \boldsymbol{y}_{k,t}^{M}~;
    &\forall k, t \in \mathscr{K}, \mathscr{T} \label{Uncertainty: Asymetric_DAprice}
\end{align}
\end{subequations}
\endgroup

We next define the uncertainty entering the constraints. The uncertain quantities comprise \ac{ndrs} availability and demand. For each period $t$ and bound $k$, $\tilde{\boldsymbol{P}}_{u,t}$ contains their nominal values, while $\check{\boldsymbol{P}}_{u,k,t}$ and $\hat{\boldsymbol{P}}_{u,k,t}$ contain the corresponding adverse deviations: reduced availability for \ac{ndrs} and increased consumption for demand, respectively. The diagonal matrix $\mathbf{S}$ assigns a positive sign to the \ac{ndrs} component and a negative sign to the demand component, allowing their uncertainty limits to be expressed using a common less-than-or-equal relation.

\begin{equation*}
\tilde{\boldsymbol{P}}_{u,t}\!\triangleq\!
\begin{bmatrix}\tilde{P}_{r,t}\\[0.2mm] \tilde{P}_{d,t}\end{bmatrix}\!,\;\;
\check{\boldsymbol{P}}_{u,k,t}\!\triangleq\!
\begin{bmatrix}\check{P}_{r,k,t}\\[0.2mm] 0\end{bmatrix}\!,\;\;
\hat{\boldsymbol{P}}_{u,k,t}\!\triangleq\!
\begin{bmatrix}0\\[0.2mm] \hat{P}_{d,k,t}\end{bmatrix}\!,\;\;
\mathbf{S}\!\triangleq\!
\begin{bmatrix}1&0\\[0.2mm]0&-1\end{bmatrix}\!,
\end{equation*}

The remaining outer-level constraints are collected in~\eqref{Uncertainty_ALL}. The matrix $\mathbf{G}$ and vector $\boldsymbol{h}$ collect the coefficients of the constraints affected by uncertainty, whereas the matrix $\mathbf{A}$ and vector $\boldsymbol{b}$ collect the coefficients of the deterministic outer-level constraints. Constraint~\eqref{Uncertainty: NDRES} gives the robust counterpart associated with \ac{ndrs} availability and demand uncertainty. The vector $\boldsymbol{\Theta}_{u,t}$, obtained from the inner uncertainty-protection problem, represents the corresponding worst-case adverse deviations. Constraint~\eqref{Uncertainty: Other_Cons_Det} retains all deterministic constraints that do not depend on uncertain parameters.
\begingroup
\allowdisplaybreaks
\begin{subequations}
\begin{align}
    & \mathbf{G}\,\mathbf{x} + \boldsymbol{h} \le \mathbf{S}\tilde{\boldsymbol{P}}_{u,t} - \boldsymbol{\Theta}_{u,t}~; %\mathop{\max}\limits_{\boldsymbol{z}_{u,k,t}\in \mathscr{Z}_{u}} \sum\limits_{k \in \mathscr{K}} \big(\check{\boldsymbol{P}}_{u,k,t} + \hat{\boldsymbol{P}}_{u,k,t}\big)\odot \boldsymbol{z}_{u,k,t}~;
    & \forall u,t \label{Uncertainty: NDRES} \\
    & \mathbf{A}\,\mathbf{x} + \boldsymbol{b} \le \mathbf{0}~;
    &
    \label{Uncertainty: Other_Cons_Det}
\end{align}
\label{Uncertainty_ALL}
\end{subequations}
\endgroup

\subsubsection{Inner Problems Reformulation}
\label{subsubsec:Inner_Reformulation}

All vector maximizations and inequalities in the following formulation are interpreted componentwise. For fixed outer-level variables $\boldsymbol{y}_{k,t}^{M}$, the robust cost $\boldsymbol{\Theta}^{M}$~\cite{morales2018robust} is obtained from the linear program~\eqref{Protection_Function_obj}. The selection vector $\boldsymbol{z}_{k,t}^{M}\triangleq[z_{k,t}^{DA},z_{k,t}^{SR,\uparrow},z_{k,t}^{SR,\downarrow}]^{\top}\!\ge\!\mathbf{0}$ indicates which market products are assigned to bound $k$ in period $t$. The objective~\eqref{Protection_obj: obj} accumulates the resulting losses due to market-price uncertainty. For each bound and product,~\eqref{Protection_obj: con1} limits the total number of selected periods according to the uncertainty-budget vector $\boldsymbol{\Gamma}_k^{M}\!\triangleq\![\Gamma_k^{DA},\Gamma_k^{SR,\uparrow},\Gamma_k^{SR,\downarrow}]^{\top}$; its dual vector is $\boldsymbol{\phi}_k^{M}\!\triangleq\![\phi_k^{DA},\phi_k^{SR,\uparrow},\phi_k^{SR,\downarrow}]^{\top}$. Constraint~\eqref{Protection_obj: con4} allows each product in a given period to be assigned to at most one bound; its dual vector is $\boldsymbol{\zeta}_t^{M}\!\triangleq\![\zeta_t^{DA},\zeta_t^{SR,\uparrow},\zeta_t^{SR,\downarrow}]^{\top}$.

\begingroup
\allowdisplaybreaks
\begin{subequations}
\begin{align}
    \boldsymbol{\Theta}^{M} = \max_{\boldsymbol{z}_{k,t}^{M}\ge \mathbf{0}}\; & \sum_{t \in \mathscr{T}}\sum_{k \in \mathscr{K}} \check{\boldsymbol{\lambda}}_{k,t}^{M}\odot \boldsymbol{y}_{k,t}^{M}\odot\boldsymbol{z}_{k,t}^{M}
     \label{Protection_obj: obj} \\
    \text{s.t.}\;\; & \sum_{t \in \mathscr{T}} \boldsymbol{z}_{k,t}^{M} \le \boldsymbol{\Gamma}_k^{M}: \boldsymbol{\phi}_k^{M}~;
    & \forall k \in \mathscr{K} \label{Protection_obj: con1} \\
    & \sum_{k \in \mathscr{K}} \boldsymbol{z}_{k,t}^{M} \le \mathbf{1}: \boldsymbol{\zeta}_t^{M}~;
    & \forall t \in \mathscr{T} \label{Protection_obj: con4}
\end{align}
\label{Protection_Function_obj}
\end{subequations}
\endgroup

The inner linear problem that determines the worst-case adverse deviations in \ac{ndrs} availability and demand, represented by $\boldsymbol{\Theta}_{u,t}$ in~\eqref{Uncertainty: NDRES}, is written explicitly in~\eqref{Protection_Function_NDRES}. For a given uncertain unit and period, the objective~\eqref{Protection_NDRES: obj} selects the bound that yields the largest adverse deviation. Constraint~\eqref{Protection_NDRES: con1} enforces the bound-specific budget $\boldsymbol{\Gamma}_{u,k}$ by accounting for the optimal selections $\boldsymbol{z}_{u,k,t^\prime}^{*}$ assigned to the other periods; its associated dual vector is $\boldsymbol{\phi}_{u,k}$. Constraint~\eqref{Protection_NDRES: con2} permits at most one active bound in period $t$; its associated dual vector is $\boldsymbol{\zeta}_{u,t}$.

\begingroup
\allowdisplaybreaks
\begin{subequations}
\begin{align}
    \boldsymbol{\Theta}_{u,t} = \max_{\boldsymbol{z}_{u,k,t}\ge \mathbf{0}}\; & \sum_{k \in \mathscr{K}} \big(\check{\boldsymbol{P}}_{u,k,t}+\hat{\boldsymbol{P}}_{u,k,t}\big)\odot \boldsymbol{z}_{u,k,t} & \forall u,t \in \mathscr{U},\mathscr{T}
    \label{Protection_NDRES: obj} \\
    \text{s.t.}\;\; & \boldsymbol{z}_{u,k,t} + \sum_{t^\prime \neq t} \boldsymbol{z}_{u,k,t^\prime}^{*} \le \boldsymbol{\Gamma}_{u,k}: \boldsymbol{\phi}_{u,k}~;
    & \forall k \in \mathscr{K} \label{Protection_NDRES: con1} \\
    & \sum_{k \in \mathscr{K}} \boldsymbol{z}_{u,k,t} \le \mathbf{1}: \boldsymbol{\zeta}_{u,t}~;
    & \label{Protection_NDRES: con2}
\end{align}
\label{Protection_Function_NDRES}
\end{subequations}
\endgroup

\subsubsection{MILP Formulation}
\label{subsubsec:MILP_Formulation}

Applying strong duality to~\eqref{Protection_Function_obj} and~\eqref{Protection_Function_NDRES} and substituting the resulting dual expressions into the bi-level model yields the single-level \ac{milp} in~\eqref{Single-level_MILP}~\cite{morales2018robust, nemati2026multi}. The dual, binary, and auxiliary variables introduced by reformulating the inner-level problems are collected in
$\Xi^{I}\!\triangleq\!
\{\boldsymbol{\phi}_k^{M},\boldsymbol{\zeta}_t^{M},
\boldsymbol{\phi}_{u,k},\boldsymbol{\zeta}_{u,t},
\boldsymbol{\chi}_{u,k,t},\boldsymbol{y}_{u,k,t}\}$. In the objective~\eqref{MILP: Obj}, the nominal profit is reduced by the dual representation of the market-price robust cost $\mathbf{1}^{\top}\boldsymbol{\Theta}^{M}$. Constraint~\eqref{MILP_asymetric1} is identical to~\eqref{Uncertainty: Asymetric_DAprice}, and~\eqref{MILP: DAMprice} imposes the dual-feasibility conditions associated with market-price uncertainty.

The asset-side robust counterpart is represented by~\eqref{MILP: NDRES1}--\eqref{MILP: Demand3}. Constraint~\eqref{MILP: NDRES1} adjusts the nominal uncertain limits using the robust protection term. Because the residual budget in~\eqref{Protection_NDRES: con1}, given by $\boldsymbol{\Gamma}_{u,k} - \sum_{t^\prime \neq t} \boldsymbol{z}_{u,k,t^\prime}^{*}$, can take only the values 0 or 1, the binary vector $\boldsymbol{\chi}_{u,k,t}$ is introduced to represent this logical condition. The nonnegative auxiliary vector $\boldsymbol{y}_{u,k,t}$ is then defined in~\eqref{MILP: NDRES1} and constrained by the Big-M inequalities in~\eqref{MILP: NDRES2}, thereby linearizing the product $\boldsymbol{\chi}_{u,k,t}\odot(\boldsymbol{\phi}_{u,k}+\boldsymbol{\zeta}_{u,t})$ that arises in the dual reformulation. Constraint~\eqref{MILP: NDRES4} allows at most one active bound per period, whereas~\eqref{MILP: NDRES5} assigns to bound $k$ exactly the number of periods prescribed by $\boldsymbol{\Gamma}_{u,k}$. The dual-feasibility conditions for reduced \ac{ndrs} availability and increased demand are given by~\eqref{MILP: NDRES3} and~\eqref{MILP: Demand3}, respectively. The domains of the auxiliary and binary variables are specified in~\eqref{MILP: Auxilliary_obj}--\eqref{MILP: Binary_cons}, and~\eqref{MILP: Other_Cons_Det} retains the deterministic constraints unaffected by uncertainty.
\vspace{2em}

\begingroup
\begin{subequations}
\allowdisplaybreaks
\begin{align} \label{MILP: Obj}
&\mathop{\max}\limits_{\Xi^{O}, \Xi^{I}}\;
\sum\limits_{t \in \mathscr{T}} \big(\boldsymbol{q}_t^{M}\big)^{\!\top}\tilde{\boldsymbol{\lambda}}_t^{M}
\;-\; \boldsymbol{c}\,\mathbf{x}
\;-\; \sum\limits_{k \in \mathscr{K}} \big(\boldsymbol{\Gamma}_k^{M}\big)^{\!\top}\boldsymbol{\phi}_k^{M} \;-\; \sum\limits_{t \in \mathscr{T}} \mathbf{1}^{\top}\boldsymbol{\zeta}_t^{M}
\end{align}
\begin{align}
    \nonumber\text{s.t.} \\
    &-\boldsymbol{\rho}_{k,t}^{M}\odot \boldsymbol{y}_{k,t}^{M} \le \boldsymbol{e}_{t}^{M} \le \boldsymbol{y}_{k,t}^{M}~;
    & \forall k, t \in \mathscr{K}, \mathscr{T} \label{MILP_asymetric1} \\
    &\boldsymbol{\phi}_k^{M} + \boldsymbol{\zeta}_t^{M} \ge  \check{\boldsymbol{\lambda}}_{k,t}^{M}\odot \boldsymbol{y}_{k,t}^{M}~;
    & \forall k, t \in \mathscr{K}, \mathscr{T} \label{MILP: DAMprice} \\
    & \mathbf{G}\,\mathbf{x} + \boldsymbol{h} \le \mathbf{S}\tilde{\boldsymbol{P}}_{u,t} - \sum\limits_{k \in \mathscr{K}} \boldsymbol{y}_{u,k,t}~;
    & \forall u,t \label{MILP: NDRES1} \\
    &\boldsymbol{\phi}_{u,k} + \boldsymbol{\zeta}_{u,t} - M(\mathbf{1} - \boldsymbol{\chi}_{u,k,t}) \le \boldsymbol{y}_{u,k,t} \le \boldsymbol{\phi}_{u,k} + \boldsymbol{\zeta}_{u,t} +M\,\boldsymbol{\chi}_{u,k,t}~;
    & \forall u,k,t \label{MILP: NDRES2} \\
    & \sum_{k\in \mathscr{K}} \boldsymbol{\chi}_{u,k,t} \le \mathbf{1}~;
    & \forall u,t \label{MILP: NDRES4} \\
    & \sum_{t\in \mathscr{T}} \boldsymbol{\chi}_{u,k,t} = \boldsymbol{\Gamma}_{u,k}~;
    & \forall u,k \label{MILP: NDRES5} \\
    &\boldsymbol{\phi}_{u,k} + \boldsymbol{\zeta}_{u,t} \ge  \check{\boldsymbol{P}}_{u,k,t}~;
    & \forall u,k,t \label{MILP: NDRES3} \\
    &\boldsymbol{\phi}_{u,k} + \boldsymbol{\zeta}_{u,t} \ge  \hat{\boldsymbol{P}}_{u,k,t}~;
    & \forall u,k,t \label{MILP: Demand3} \\
    &\boldsymbol{\phi}_k^{M}, \boldsymbol{\zeta}_t^{M} %,\boldsymbol{y}_{k,t}^{M}
    \ge \mathbf{0}~;
    & \forall k,t \in \mathscr{K},\mathscr{T} \label{MILP: Auxilliary_obj} \\
    & \boldsymbol{\phi}_{u,k}, \boldsymbol{\zeta}_{u,t}, \boldsymbol{y}_{u,k,t} \ge \mathbf{0}~;
    & \forall u,k,t \label{MILP: Auxilliary_cons} \\
    & \boldsymbol{\chi}_{u,k,t} \in \{0,1\}^{2}~;
    & \forall u,k,t \label{MILP: Binary_cons} \\
    %& \chi_{r,k,t}, \chi_{d,k,t} \in \{0, 1\}~; 
    %\hspace{16.2em}   
    %\forall r,d,k, t \in \mathscr{R},\mathscr{D},\mathscr{K},\mathscr{T} 
    %
    & \mathbf{A}\,\mathbf{x} + \boldsymbol{b} \le \mathbf{0}~;
    &
    \label{MILP: Other_Cons_Det}
\end{align}
\label{Single-level_MILP}
\end{subequations}
\endgroup

\vspace{-1em}
\subsection{Probability-Layer Clustering and Data-Driven Bound-and-Budget Calibration}
\label{subsec:Calibration_MB}

The \ac{plmbro} formulation developed in~\Cref{subsec:Multi-Bound_Robust} requires nominal reference profiles, deviation bounds, and full-horizon uncertainty budgets for each uncertain parameter. This subsection describes how these quantities are calibrated from historical data. Unlike the percentile-based \ac{mbro} benchmark~\cite{nemati2026multi}, which uses predefined percentile bounds and externally selected budget vectors, the proposed procedure constructs parameter-specific probability layers and determines their associated budgets from the empirical distribution of the historical profiles.
\\

\noindent\textbf{Nominal Reference Profiles
}

\noindent For a generic uncertain parameter, let $\boldsymbol{\omega}^{(n)}\!\triangleq\![\omega_{1}^{(n)},\ldots,\omega_{H}^{(n)}]^{\top}$ denote historical daily profile $n$, where $n=1,\ldots,N$ and $H\!\triangleq\!|\mathscr{T}|$ is the number of periods in the scheduling horizon. The nominal reference profile is defined as the coordinate-wise median of the complete historical sample:
\begin{equation*}
\tilde{\omega}_{t}\!\triangleq\!
\operatorname{median}\big\{\omega_{t}^{(1)},\ldots,\omega_{t}^{(N)}\big\},
\qquad \forall t\in\mathscr{T}.
\end{equation*}
The median reference is used for all uncertain parameters and is consistent with the Manhattan ($\ell_1$) distance criterion adopted in the clustering procedure. Nevertheless, the proposed calibration is not restricted to this choice: an external point forecast or a forecast produced by another statistical method can replace $\tilde{\omega}_{t}$ without modifying the clustering or budget-calibration steps. In that case, the deviation bounds are simply measured relative to the alternative reference profile.
\\

\noindent\textbf{Full-horizon Uncertainty Budgets}

\noindent 
The historical profiles of each uncertain parameter are clustered separately using the $k$-medians algorithm with the Manhattan distance. In generic form, the clustering problem is
\begin{equation*}
\min_{\{\mathscr{I}_{k},\boldsymbol{\mu}_{k}\}_{k\in\mathscr{K}}}
\sum_{k\in\mathscr{K}}
\sum_{n\in\mathscr{I}_{k}}
\left\|
\boldsymbol{\omega}^{(n)}-\boldsymbol{\mu}_{k}
\right\|_{1},
\qquad
\mu_{k,t}\!\triangleq\!
\operatorname{median}
\big\{\omega_{t}^{(n)}:n\in\mathscr{I}_{k}\big\},
\end{equation*}
where $\mathscr{I}_{k}$ is the index set of historical profiles assigned to cluster $k$, and $\boldsymbol{\mu}_{k}$ is the corresponding coordinate-wise median profile. The sets $\mathscr{I}_{k}$ form a nonempty partition of $\{1,\ldots,N\}$. The Manhattan distance and coordinate-wise median centers reduce the influence of isolated extreme observations relative to squared-error clustering criteria. Because the clustering problem can admit different local solutions depending on initialization, the implementation retains the solution with the smallest within-cluster $\ell_{1}$ objective among the initializations considered.

The number of clusters $|\mathscr{K}|$ determines the number of probability layers in the multi-bound uncertainty set. The proposed procedure is applicable to any number of clusters, which may be selected according to clustering-validity criteria, the desired level of representation, and the computational size of the resulting robust model. The numerical study uses three clusters as a practical compromise between the representation of historical variability and computational complexity, while also permitting direct comparison with the three-bound percentile-based benchmark. Once the clusters are obtained, they may be indexed in ascending order of the temporal averages of their median profiles.

The empirical probability of cluster $k$ is
\begin{equation*}
\Pi_{k}\!\triangleq\!\frac{|\mathscr{I}_{k}|}{N},
\qquad
\sum_{k\in\mathscr{K}}\Pi_{k}=1.
\end{equation*}
The corresponding full-horizon budgets are obtained by mapping the empirical probabilities to the $H$ scheduling periods. The integer budget vector is allocated using the largest-remainder method:
\begin{equation*}
\big(\Gamma_{k}\big)_{k\in\mathscr{K}}
=
\operatorname{LR}\!\left(
\big(H\Pi_{k}\big)_{k\in\mathscr{K}}
\right),
\qquad
\sum_{k\in\mathscr{K}}\Gamma_{k}=H,
\end{equation*}
where $\operatorname{LR}(\cdot)$ denotes the largest-remainder allocation: the values $H\Pi_k$ are first rounded downward, and the remaining periods are assigned to the layers with the largest fractional remainders until the total equals $H$. Thus, $\Gamma_k$ is the integer full-horizon budget associated with probability layer $k$. The mapping transfers the empirical prevalence of historical daily-profile regimes to horizon-wide shares of the robust deviation layers; it should therefore be interpreted as a calibration rule rather than as a direct estimate of period-level deviation probabilities. The symbols $\Pi_k$ and $\Gamma_k$ are used generically in this subsection; because clustering is performed separately for every uncertain parameter, the resulting budgets populate the corresponding components of $\boldsymbol{\Gamma}_{k}^{M}$ and $\boldsymbol{\Gamma}_{u,k}$ in the robust formulation.
\\

\vspace{4em}
\noindent\textbf{Deviation Bounds}

\noindent Each cluster also supplies strategy-dependent representative profiles from which its deviation bounds are calculated. For one-sided uncertain parameters, the balanced strategy uses the cluster median $\boldsymbol{\mu}_{k}$, whereas the optimistic and pessimistic strategies use, respectively, selected favorable and adverse coordinate-wise within-cluster quantile profiles. The quantile direction depends on the adverse direction of the uncertain parameter: lower realizations are adverse for \ac{ndrs} availability and \ac{srm} prices, whereas higher realizations are adverse for demand. For the two-sided \ac{dam} price, lower- and upper-side representative profiles are constructed separately within each cluster because either a price decrease or a price increase can be adverse depending on the sign of the \ac{rvpp} energy position. The strategy-specific quantile levels are risk-attitude parameters; once they are specified, the corresponding representative profiles and deviation magnitudes are generated automatically from the historical data. Changing these levels modifies the severity of the deviation bounds but leaves the cluster assignments, empirical probabilities, and full-horizon budgets unchanged. Hence, the empirical data determine the probability layers and their budgets, while the selected risk-attitude quantiles determine the severity of the strategy-dependent representatives; conditional on these quantile levels, the resulting deviation bounds are generated automatically from the historical data.

For a selected strategy and probability layer, let $\omega_{k,t}^{\mathrm{rep},-}$ and $\omega_{k,t}^{\mathrm{rep},+}$ denote the lower- and upper-side representative values, respectively. The corresponding deviation magnitudes are

\begin{equation*}
\check{\omega}_{k,t}\!\triangleq\!
\max\big\{\tilde{\omega}_{t}-\omega_{k,t}^{\mathrm{rep},-},0\big\},
\qquad
\hat{\omega}_{k,t}\!\triangleq\!
\max\big\{\omega_{k,t}^{\mathrm{rep},+}-\tilde{\omega}_{t},0\big\}.
\end{equation*}

For one-sided uncertain parameters, only the representative associated with the adverse direction is required and the opposite deviation magnitude is set to zero. For the \ac{dam} price, both sides are retained because the adverse direction depends on whether the \ac{rvpp} sells or purchases energy. The maximum operators ensure that both deviations are nonnegative and that a representative value lying on the opposite side of the nominal reference contributes zero to the corresponding deviation magnitude.

Following the sign conventions introduced in~\Cref{subsubsec:Bilevel_formulation}, only negative deviations are retained for \ac{ndrs} availability and upward/downward \ac{srm} prices, while only positive deviations are retained for demand. For the \ac{dam} price, both lower- and upper-side deviations are retained. Accordingly, the calibration provides $\check{\boldsymbol{\lambda}}_{k,t}^{M}$, $\hat{\boldsymbol{\lambda}}_{k,t}^{M}$, $\check{\boldsymbol{P}}_{u,k,t}$, and $\hat{\boldsymbol{P}}_{u,k,t}$ for the robust formulation.

\begin{figure*}[t!]
\centering
\resizebox{\textwidth}{!}{%
\begin{tikzpicture}[
    font=\small,
    >={Stealth[length=2.2mm]},
    blackbox/.style={draw, dashed, rounded corners, fill=gray!5, inner sep=6pt, align=left},
    eqbox/.style={draw, rounded corners, fill=gray!3, inner sep=8pt, align=left},
]
\colorlet{clu}{orange!80!black}

% ===================== (a) detailed DAM clusters + black box for the others =====================
\node[anchor=north west, font=\bfseries] (TA) at (0,0)
  {(a) Data-driven bounds and budgets obtained by clustering the historical daily profiles};

\node[anchor=north west, inner sep=0pt] (A) at (0,-0.5)
  {\includegraphics[width=7.5cm]{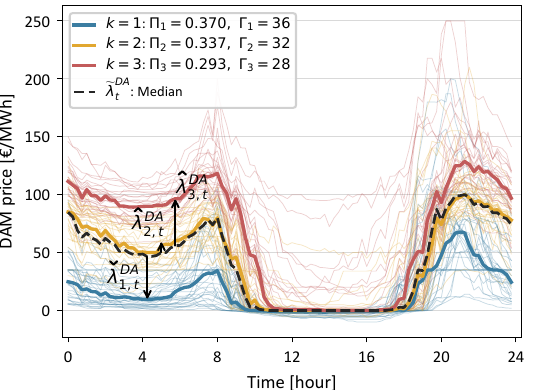}};

\node[blackbox, anchor=north west, text width=7.9cm] (OTH) at (7.6,-0.6)
  {{\bfseries Other uncertain parameters}\\[3pt]
   \ac{ndrs} generation, demand, and upward and downward \ac{srm} prices are clustered in the \emph{same} way; each parameter yields its own:\\[3pt]
\hspace*{1em}$\bullet$~Reference profile $\tilde{\omega}_{t}$, mapped to the corresponding component of $\tilde{\boldsymbol{\lambda}}_t^{M}$ or $\tilde{\boldsymbol{P}}_{u,t}$;\\[1pt]
\hspace*{1em}$\bullet$~Cluster deviations $\check{\omega}_{k,t},\hat{\omega}_{k,t}$, mapped to the corresponding components of $\check{\boldsymbol{\lambda}}_{k,t}^{M},\hat{\boldsymbol{\lambda}}_{k,t}^{M}$ or $\check{\boldsymbol{P}}_{u,k,t},\hat{\boldsymbol{P}}_{u,k,t}$;\\[1pt]
\hspace*{1em}$\bullet$~Cluster probability $\Pi_k$;\\[1pt]
\hspace*{1em}$\bullet$~Uncertainty budget $\Gamma_k$, mapped to the corresponding component of $\boldsymbol{\Gamma}_{k}^{M}$ or $\boldsymbol{\Gamma}_{u,k}$.
   };

% ===================== bridge =====================
\node[] (BR) at (8.0,-6.7)
  {{\color{black}Cluster representatives determine the deviations \textcolor{clu}{$\check{\boldsymbol{\lambda}}_{k,t}^{M}$},\ \textcolor{clu}{$\hat{\boldsymbol{\lambda}}_{k,t}^{M}$}, \textcolor{clu}{$\check{\boldsymbol{P}}_{u,k,t}$},\ \textcolor{clu}{$\hat{\boldsymbol{P}}_{u,k,t}$}, while cluster probabilities determine the budgets \textcolor{clu}{$\boldsymbol{\Gamma}_{k}^{M}$} and \textcolor{clu}{$\boldsymbol{\Gamma}_{u,k}$}}};
  \node[] (BRarrow) at (8.0,-7.1)
  {$\Downarrow$};

% ===================== (b) coupling to the robust model =====================
\node[anchor=north west, font=\bfseries] (TB) at (0,-7.5)
  {(b) How the clustered bounds and budgets enter the robust multi-bound model};

\node[eqbox, anchor=north west, text width=15.4cm] (EQ) at (0,-7.95)
{%
{\bfseries Market-price uncertainty}\hfill{\scriptsize Eqs.~\eqref{Protection_Function_obj},~\eqref{MILP: Obj}--\eqref{MILP: DAMprice}}\\[3pt]
$\boldsymbol{\phi}_k^{M}+\boldsymbol{\zeta}_t^{M}\ \ge\ \textcolor{clu}{\check{\boldsymbol{\lambda}}_{k,t}^{M}}\odot\boldsymbol{y}_{k,t}^{M}$\hfill{\scriptsize Worst-case price loss uses the \textcolor{clu}{cluster deviation} (the positive \ac{dam} deviation enters through $\boldsymbol{\rho}_{k,t}^{M}$ in~\eqref{MILP_asymetric1})}\\[2pt]
$\textstyle\sum_{t}\boldsymbol{z}_{k,t}^{M}\le\textcolor{clu}{\boldsymbol{\Gamma}_{k}^{M}}$
\hfill{\scriptsize The \textcolor{clu}{full-horizon budgets} are obtained from the parameter-specific cluster probabilities}\\[5pt]
\rule{\linewidth}{0.3pt}\\[4pt]
{\bfseries Unit uncertainty (\ac{ndrs} and demand)}\hfill{\scriptsize Eqs.~\eqref{Protection_Function_NDRES},~\eqref{MILP: NDRES1}--\eqref{MILP: Demand3}}\\[3pt]
Calibrated separately using the same procedure: negative deviations are retained for \ac{ndrs} availability, whereas positive deviations are retained for demand. The resulting \textcolor{clu}{$\check{\boldsymbol{P}}_{u,k,t}$} and \textcolor{clu}{$\hat{\boldsymbol{P}}_{u,k,t}$}, together with the budgets \textcolor{clu}{$\boldsymbol{\Gamma}_{u,k}$}, tighten the corresponding uncertainty limits.\\[5pt]
\rule{\linewidth}{0.3pt}\\[4pt]
{\bfseries Deterministic constraints}\hfill{\scriptsize Eqs.~\eqref{cons:Supply-Demand1},~\eqref{Deterministic: DRES},~\eqref{cons: NDRES2},~\eqref{cons: Demand2}--\eqref{cons: Demand3},~\eqref{Deterministic: ES},~\eqref{Deterministic: Reserve}}\\[3pt]
Unaffected by uncertainty; gathered compactly as $\mathbf{A}\mathbf{x}+\boldsymbol{b}\le\mathbf{0}$.%
};

\end{tikzpicture}%
}
\caption{Probability-layer construction of the multi-bound uncertainty set and its connection to the robust formulation. Panel~(a) illustrates the three-cluster configuration adopted in the numerical study for the \ac{dam} price. The coordinate-wise median of the complete historical sample defines the nominal reference profile, while the strategy-dependent cluster representatives determine the deviation layers and the empirical cluster probabilities determine their full-horizon budgets through the largest-remainder allocation. The same parameter-specific procedure is applied separately to the upward and downward \ac{srm} prices, \ac{ndrs} availability, and demand, following their corresponding adverse-deviation directions. Panel~(b) shows representative locations at which the calibrated deviations and budgets highlighted in color enter the market-price and unit-uncertainty terms of the robust formulation. The procedure is applicable to any number of clusters and scheduling periods; the three-cluster configuration shown here corresponds to the numerical implementation.}
\label{fig:MBRO_Procedure}
\end{figure*}

Figure~\ref{fig:MBRO_Procedure} summarizes the connection between the probability-layer calibration and the robust model. Panel~(a) illustrates the construction of the reference profile, cluster representatives, empirical probabilities, and full-horizon budgets. The three-cluster example corresponds to the configuration adopted in the numerical study, whereas the procedure itself is valid for any $|\mathscr{K}|$. Panel~(b) identifies, in color, the calibrated deviation bounds and budgets that enter the market-price and unit-uncertainty expressions in~\Cref{subsec:Multi-Bound_Robust}. The deterministic constraints remain unaffected by this calibration. Once these inputs are constructed, the robust inner problems retain their adversarial role of selecting the periods in which the available deviation layers are most harmful to the \ac{rvpp} bidding decisions.

The numerical data, selected number of clusters, strategy-specific within-cluster quantile levels, and resulting parameter-specific budgets are reported in~\Cref{sec:Case_Studies}.

% ---------------- Figure ----------------

\clearpage

\section{Case Studies}
\label{sec:Case_Studies}

This section evaluates the proposed \ac{plmbro} framework for the joint participation of an \ac{rvpp} in the \ac{dam} and \ac{srm} under market price, renewable generation, and demand consumption uncertainties and compares its economic performance with the percentile-based \ac{mbro} benchmark. Special emphasis is placed on the data-driven calibration of parameter-specific deviation bounds and uncertainty budgets and on the resulting in-sample and out-of-sample economic performance. All optimization models were implemented in GAMS~39.1.1 and solved with CPLEX on a Dell XPS laptop equipped with an Intel i7-1165G7 processor (2.8~GHz) and 16~GB of RAM.

\subsection{System Description and \ac{plmbro} Calibration}

The proposed framework is evaluated on a representative \ac{rvpp} located in Spain that aggregates a hydro plant, a wind farm, a solar \ac{pv} plant, a Li-ion \ac{ess}, and a flexible demand. The portfolio is deliberately conceived as a balanced test system rather than a replica of a single physical plant: the solar \ac{pv} plant, the wind farm, and the flexible demand share the same 50~MW rating, so that the comparison between the proposed and the benchmark uncertainty-handling models is not dominated by any single technology, while the distinctive behavior of each asset is reproduced through its generation/consumption profile, operating limits, and uncertainty description. The uncertain inputs---wind and solar \ac{pv} generation, demand, \ac{dam} prices, and upward and downward \ac{srm} prices---are obtained from quarter-hourly Spanish historical records reported in~\cite{REE2025}. The data used in the numerical assessment comprise 92 calibration days and 20 held-out days from the same seasonal regime; the former are used to construct the uncertainty set and for the in-sample assessment, whereas the latter are reserved exclusively for the out-of-sample assessment. The methodology is not tied to this particular data window and can be applied to longer historical records as additional quarter-hourly observations become available. For every uncertain parameter, the calibration days are partitioned into three representative clusters ($k=1,2,3$). Once the strategy-specific quantile levels are specified, the cluster representatives determine the deviation bounds and the empirical cluster probabilities determine the full-horizon uncertainty budgets through the largest-remainder allocation described in~\Cref{subsec:Calibration_MB}.

For the proposed \ac{plmbro} model, the balanced representative of each cluster is based on the cluster median. For the one-sided uncertain parameters, the 20th and 80th within-cluster percentiles are used to construct the optimistic and pessimistic representatives, with their assignment determined by the adverse direction of the uncertain parameter: for \ac{ndrs} availability and \ac{srm} prices, the 80th percentile is used for the optimistic strategy and the 20th percentile for the pessimistic strategy, whereas for demand the assignment is reversed. For the two-sided \ac{dam} price, the lower- and upper-side representatives are constructed separately using the corresponding strategy-dependent within-cluster quantiles.

The technical characteristics of the \ac{rvpp} assets are summarized in~\Cref{table:Data_Det} and are taken from~\cite{nemati2025segan}. Two uncertainty-handling models are compared throughout this section. The proposed \ac{plmbro} model generates the strategy-dependent deviation bounds from the historical records once the quantile levels are specified and derives the uncertainty budgets automatically from the empirical cluster probabilities, following the clustering procedure detailed in~\Cref{subsec:Calibration_MB}; the resulting per-parameter budgets are reported in~\Cref{table:Cluster_Budget}. The benchmark \ac{mbro} model instead follows the percentile-based calibration of~\cite{nemati2026multi}, in which the deviation bounds are obtained from fixed percentile pairs and the uncertainty budgets are specified externally and enlarged as the operator becomes more conservative, as listed in~\Cref{table:Data_Budget}. For the proposed \ac{plmbro} calibration, the three data-driven layer budgets sum to the 96 quarter-hourly periods for each uncertain parameter, while the at-most-one-layer constraints prevent a period from being assigned to more than one uncertainty layer. By contrast, the benchmark uses predefined, user-specified uncertainty budgets whose sum is less than or equal to 96, so not every period is necessarily assigned to a deviation layer.

\begin{figure}[t!]
    \centering
        {\includegraphics[width=.84\linewidth]{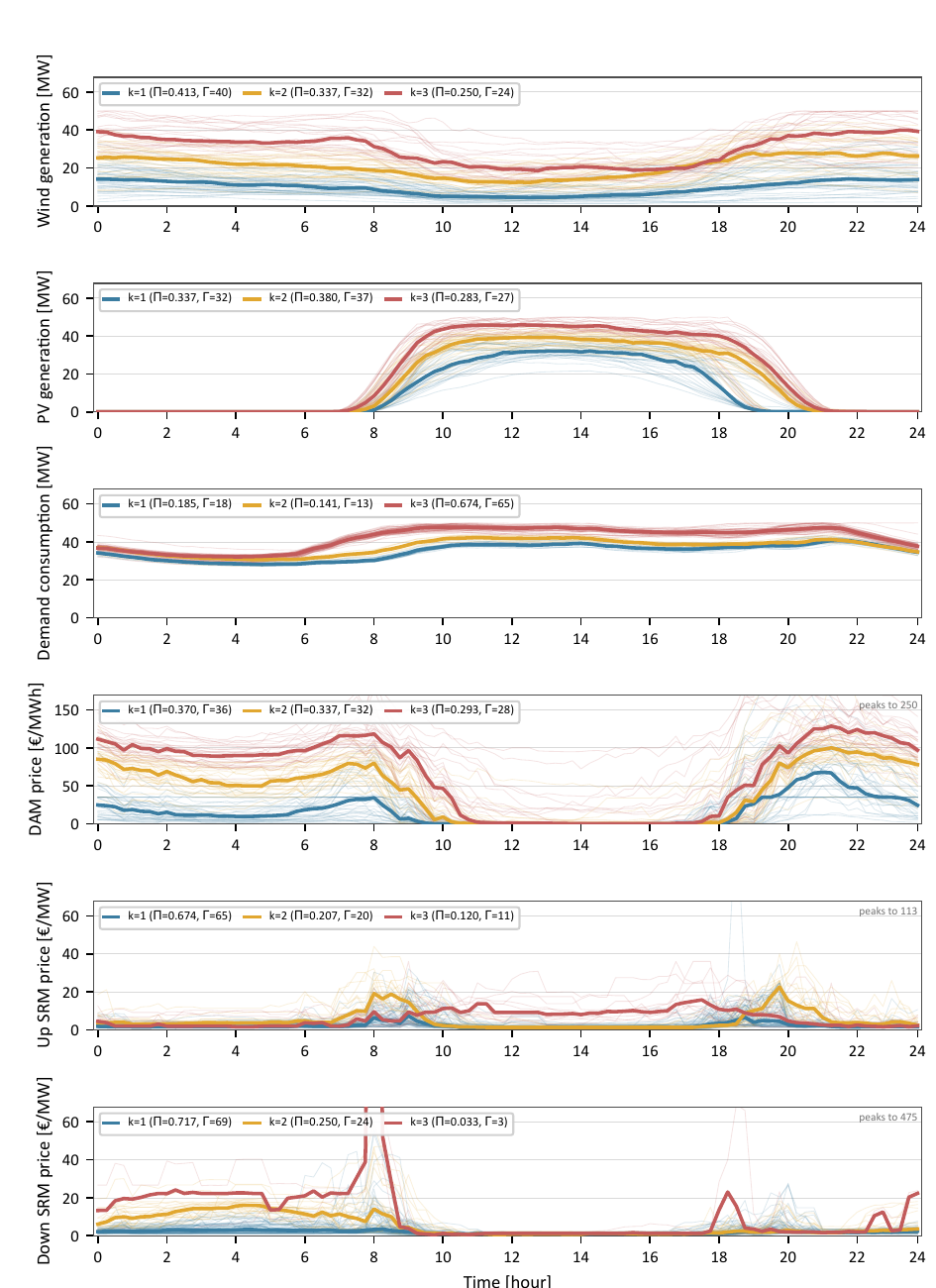}}
    \caption{Representative clusters of the uncertain inputs used to build the multi-bound uncertainty set. For each parameter, the bold curves denote the cluster medians and the light curves the historical member days assigned to each cluster; $\Pi$ is the cluster probability and $\Gamma$ the resulting uncertainty budget.}
    \label{fig:Uncertainty_Bound_data}
    %\vspace{3em}
\end{figure}

\begin{figure}[t!]
    \centering
        {\includegraphics[width=\linewidth]{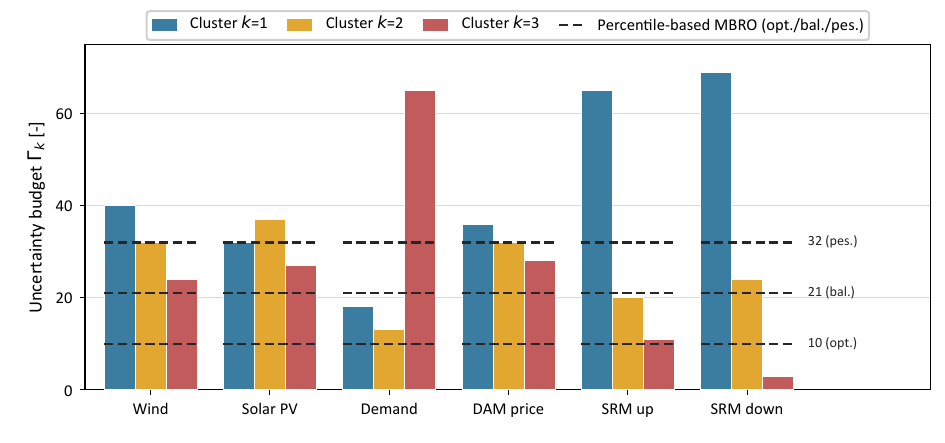}}
    \caption{Cluster-based uncertainty budgets $\Gamma_k$ obtained from the empirical cluster probabilities for each uncertain parameter, compared with the strategy-dependent uniform budgets of the percentile-based \ac{mbro} benchmark (dashed lines). The proposed budgets are parameter-specific and generally non-uniform.}
    \label{fig:Cluster_Budget_fig}
    \vspace{4em}
\end{figure}

\Cref{table:Cluster_Budget} reports the budgets obtained for the six uncertain parameters, whose clusters are depicted in~\Cref{fig:Uncertainty_Bound_data}, while \Cref{fig:Cluster_Budget_fig} contrasts their parameter-specific allocations with the uniform allocation used by the percentile-based benchmark. The empirical layer frequencies are markedly non-uniform and differ substantially across parameters. The \ac{dam} price is distributed relatively evenly across the three clusters, with $(\Gamma_1,\Gamma_2,\Gamma_3)=(36,32,28)$, whereas the upward and downward \ac{srm} prices are strongly concentrated in their lowest-level clusters, with budgets $(65,20,11)$ and $(69,24,3)$, respectively. Demand exhibits the opposite pattern, with its highest-level cluster receiving 65 of the 96 periods. Thus, the proposed calibration assigns each uncertainty layer a full-horizon budget according to its empirical cluster frequency rather than imposing the same allocation across parameters and layers. These budgets define the allowable horizon-wide allocation of each deviation layer, while the robust inner problem still determines the periods in which the permitted deviations are most adverse to the \ac{rvpp} decisions.

\begin{table}[t!]
  \centering
   % \small
\caption{Technical data of the RVPP units.}
    \label{table:Data_Det}
  %\small
  \setlength{\tabcolsep}{2pt}
  \renewcommand{\arraystretch}{1.4}
 %\small
  \begin{threeparttable}
  \vspace{-.5em}
  \begin{tabular}{lccccccccccc}
    \toprule

    \multicolumn{1}{c}{\textbf{Parameter}}  
    && \multicolumn{1}{c}{\textbf{Solar PV}}
    && \multicolumn{1}{c}{\textbf{Wind farm}} 
    && \multicolumn{1}{c}{\textbf{Hydro}}
    && \multicolumn{1}{c}{\textbf{ESS}}
    && \multicolumn{1}{c}{\textbf{Demand}}
    \\

    \cmidrule{1-1} \cmidrule{3-3} \cmidrule{5-5} \cmidrule{7-7} \cmidrule{9-9} \cmidrule{11-11} 

   \multirow{1}{*}{\text{$\bar P_{u}$/$\ubar P_{u}$ [MW]}}    && 50/0 
   && 50/0 && 50/10 && 10/10 && 50/0 \\ [0.2em]

     \multirow{1}{*}{\text{$R_{u}^{\uparrow}$/$R_{u}^{\downarrow}$ [MW/min]}}    && 20/25 && 15/20 && 10/10 && 10/10 && 3/3  \\ [0.2em]

          \multirow{1}{*}{\text{$\beta_{u}^{\uparrow}$/$\beta_{u}^{\downarrow}$ [\%]}}    && 5/5 && 5/5 && 50/50 && 100/100 && 0/0  \\ [0.2em]

        %\multirow{1}{*}{\text{$UT_u$/$DT_u$ [hour]}}    && - && - && 1/0 && - &&  - \\ [0.2em]

        \multirow{1}{*}{\text{$\bar E_{u}/\ubar E_{u}$ [MWh]}}    && - && - && 480/- && 30/3 && -/750 \\ [0.2em]

        \multirow{1}{*}{\text{$\bar P_{s}^{+}/\bar P_{s}^{-}$ [MW]}}   && - && - && - && 10/10 && - \\ [0.2em]

        \multirow{1}{*}{\text{$\eta^{+}_s/\eta^{-}_s$ [\%]}}    && - && - && - && 95/95 && - \\ [0.2em]

        \multirow{1}{*}{\text{$C_{u}$ [€/MWh]}}    && 10 && 15 && 12.5 && 30 && -  \\ [0.2em]

\bottomrule
  \end{tabular}
\end{threeparttable}
%\vspace{1mm}
\end{table}

\begin{table}[t!]
  \centering
  \caption{Externally specified uncertainty budgets of the benchmark \ac{mbro} model for the three uncertainty-handling strategies.}
  %\small
  \setlength{\tabcolsep}{9pt}
  \renewcommand{\arraystretch}{1.4}
  \vspace{-.5em}
  \begin{threeparttable}
  \begin{tabular}{lcccc}
    \toprule

    \multicolumn{1}{c}{\textbf{Strategy}}
    && \multicolumn{1}{c}{\textbf{$\Gamma_{k=1}$}}
    & \multicolumn{1}{c}{\textbf{$\Gamma_{k=2}$}}
    & \multicolumn{1}{c}{\textbf{$\Gamma_{k=3}$}}
    \\

 \cmidrule{1-1} \cmidrule{3-5}

    \multirow{1}{*}{Optimistic}  && 10 & 10 & 10 \\ [0.2em]

    \multirow{1}{*}{Balanced}    && 21 & 21 & 21 \\ [0.2em]

    \multirow{1}{*}{Pessimistic} && 32 & 32 & 32 \\ [0.2em]

\bottomrule
  \end{tabular}
  \begin{tablenotes}
  \footnotesize
  \item Within each strategy, the same externally specified budget is assigned to all three bounds and all uncertain parameters. The common budget increases from the optimistic to the pessimistic strategy to represent increasing conservatism.
  \end{tablenotes}
\end{threeparttable}
  \label{table:Data_Budget}
   % \vspace{1em}
\end{table}

\begin{table}[t!]
  \centering
  \caption{Cluster-based uncertainty budgets $\Gamma_k$ and empirical cluster probabilities $\Pi_k$ obtained for each uncertain parameter.}
  \setlength{\tabcolsep}{10pt}
  \renewcommand{\arraystretch}{1.4}
  \vspace{-.5em}
  \begin{threeparttable}
\begin{tabular}{lccc}
    \toprule
    \multicolumn{1}{c}{\textbf{Parameter}}
    & \multicolumn{1}{c}{\textbf{$k=1$}}
    & \multicolumn{1}{c}{\textbf{$k=2$}}
    & \multicolumn{1}{c}{\textbf{$k=3$}} \\
    & $\Gamma$ ($\Pi$) & $\Gamma$ ($\Pi$) & $\Gamma$ ($\Pi$) \\
    \cmidrule(lr){1-1} \cmidrule(lr){2-2} \cmidrule(lr){3-3} \cmidrule(lr){4-4}
    Wind        & 40 (0.413) & 32 (0.337) & 24 (0.250) \\
    Solar PV    & 32 (0.337) & 37 (0.380) & 27 (0.283) \\
    Demand      & 18 (0.185) & 13 (0.141) & 65 (0.674) \\
    DAM price   & 36 (0.370) & 32 (0.337) & 28 (0.293) \\
    SRM up      & 65 (0.674) & 20 (0.207) & 11 (0.120) \\
    SRM down    & 69 (0.717) & 24 (0.250) &  3 (0.033) \\
    \bottomrule
  \end{tabular}
  \begin{tablenotes}
  \footnotesize
    \item Clusters are indexed $k=1,2,3$ by increasing level of their median profiles. The empirical probability is calculated as $\Pi_k=|\mathscr{I}_k|/92$, where $|\mathscr{I}_k|$ is the number of calibration days assigned to cluster $k$. The corresponding integer budgets are obtained as $(\Gamma_k)_{k=1}^{3}=\operatorname{LR}\!\left((96\,\Pi_k)_{k=1}^{3}\right)$, ensuring $\sum_k\Gamma_k=96$. Probabilities are reported to three decimal places.
  \end{tablenotes}
  \end{threeparttable}
  \label{table:Cluster_Budget}
  \vspace{2em}
\end{table}

\newpage

Three uncertainty-handling strategies are considered for the \ac{rvpp} operator---optimistic, balanced, and pessimistic---which, in the proposed \ac{plmbro} model, correspond to increasingly adverse representative realizations within each cluster, as explained in~\Cref{subsec:Calibration_MB}. The following three case studies characterize the optimal operation obtained with the proposed approach, benchmark it economically against the percentile-based \ac{mbro} benchmark on in-sample data, and assess the practical performance of the resulting bids out of sample:

\begin{itemize}
\item \textbf{Case 1:} Analyze the optimal quarter-hourly operation of the \ac{rvpp} units, in terms of both energy dispatch and secondary-reserve provision, obtained with the proposed \ac{plmbro} model under the optimistic, balanced, and pessimistic strategies.

\item \textbf{Case 2:} Assess, on in-sample historical data, the economic performance of the proposed \ac{plmbro} approach against the percentile-based \ac{mbro} benchmark---in which the deviation bounds are constructed from predefined percentile levels and the uncertainty budgets are specified externally---across the three uncertainty-handling strategies.

\item \textbf{Case 3:} Assess the out-of-sample economic performance of the fixed market bids generated by the two approaches on held-out historical realizations, using realized profit and imbalance cost as the evaluation metrics.

\end{itemize}

\vspace{1.5em}
\subsection{Case 1: Optimal RVPP Operation}
\label{subsec:Case1}

\begin{figure}[t!]
    \centering
        {\includegraphics[width=\linewidth]{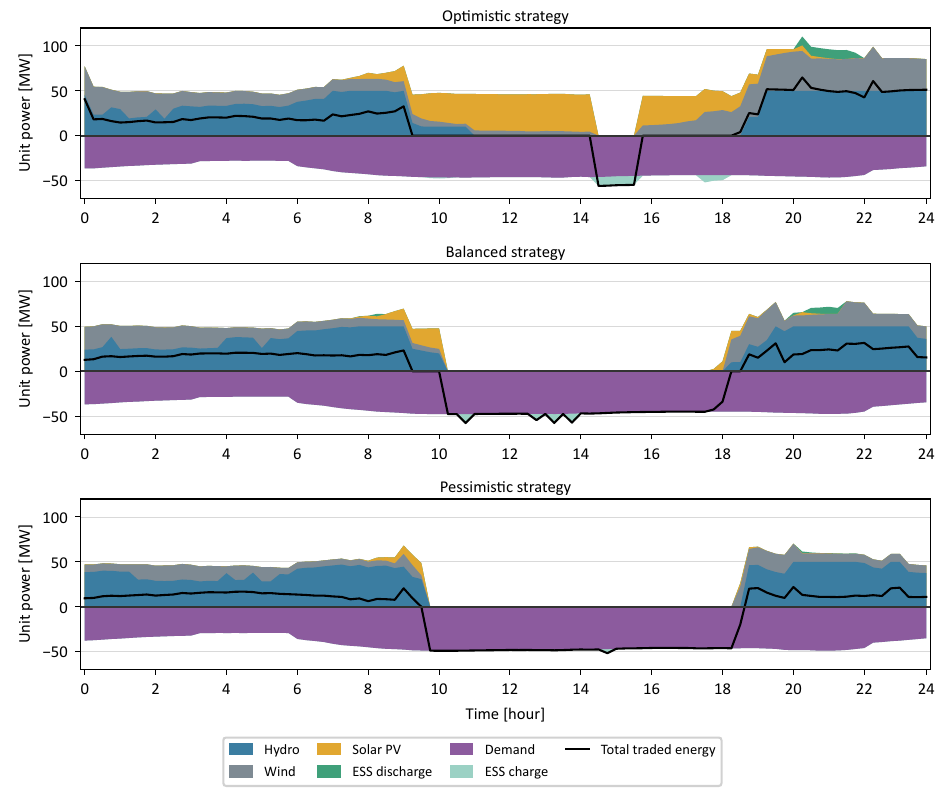}}
    \caption{Case 1 -- Quarter-hourly energy dispatch of the RVPP units under the proposed \ac{plmbro} model for the optimistic, balanced, and pessimistic strategies.}
    \label{fig:ALL_stack_Case1}
    %\vspace{1em}
\end{figure}

\Cref{fig:ALL_stack_Case1} depicts the quarter-hourly energy dispatch of the \ac{rvpp} units under the optimistic, balanced, and pessimistic strategies. As the strategy becomes more conservative, the \ac{plmbro} model uses increasingly adverse lower representatives for non-dispatchable generation and higher representatives for demand, reducing the energy volume that can be profitably committed to the \ac{dam}. This trend is reflected in the scheduled renewable output, which falls by roughly 61\% for the wind farm and by almost 95\% for the solar \ac{pv} plant between the optimistic and pessimistic strategies, while scheduled demand increases by about 4\%. The particularly pronounced reduction in solar \ac{pv} production reflects not only the lower assumed renewable availability but also the more conservative market-price realization, under which using all available solar generation becomes less economically attractive, making additional energy curtailment economically preferable. The dispatchable hydro plant remains committed under all three strategies and concentrates its generation in the high-price overnight and evening windows (approximately 00:00--10:00 and 18:45--24:00), thereby supporting market sales during the most valuable periods. The \ac{ess} complements this behavior through intertemporal arbitrage: it charges during the lower-price midday periods, when the assumed solar availability is comparatively higher, and discharges mainly during the evening price ramp (around 20:00--22:00). As conservatism increases, both the charging activity and the subsequent discharge decrease. This reduction becomes particularly pronounced under the pessimistic strategy because less renewable energy is available for charging and the more conservative \ac{dam}-price realization reduces the economic value of discharging stored energy.

\begin{figure}[t!]
    \centering
        {\includegraphics[width=\linewidth]{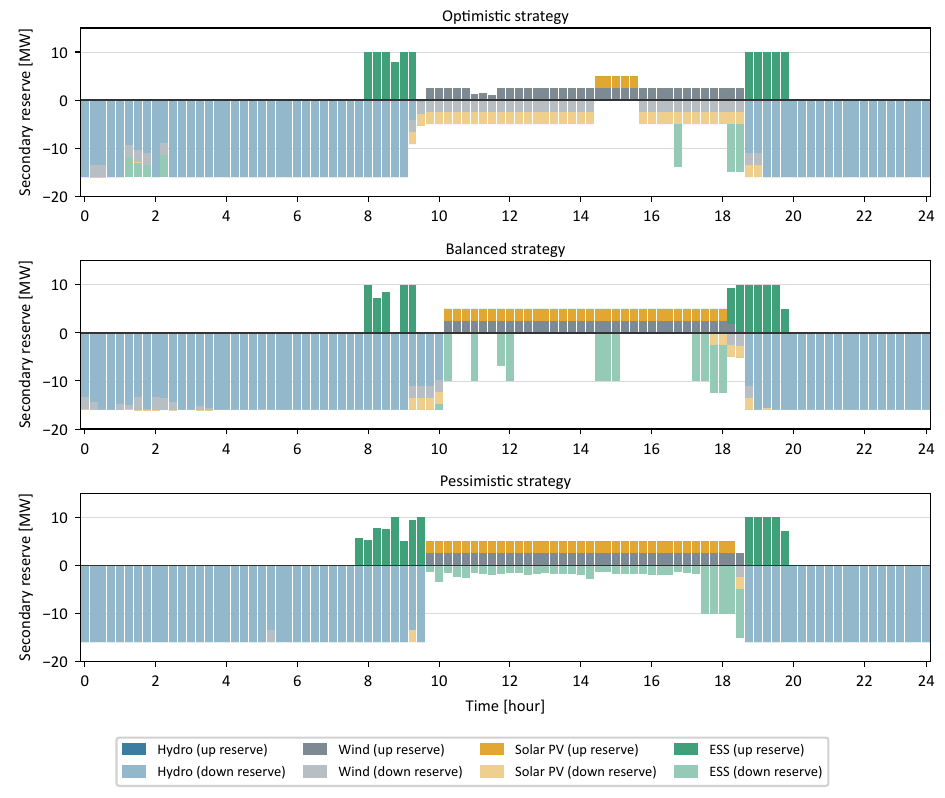}}
    \caption{Case 1 -- Upward and downward secondary-reserve provision by the RVPP units under the proposed \ac{plmbro} model for the optimistic, balanced, and pessimistic strategies.}
    \label{fig:ALL_stack_reserve_Case1}
    \vspace{-1em}
\end{figure}

\Cref{fig:ALL_stack_reserve_Case1} reports the upward and downward secondary reserve provided by each \ac{rvpp} unit. The \ac{ess} is the principal provider of upward reserve and offers it mainly in the morning (around 08:00--09:30) and the early evening (around 18:45--20:00), when upward \ac{srm} prices are most attractive. The wind and solar \ac{pv} plants also provide upward reserve during daytime periods, when available renewable generation exceeds their scheduled energy output and therefore leaves upward headroom. Downward reserve, in contrast, is dominated by the hydro plant, which can curtail its committed output during its generating windows and therefore offers downward reserve throughout most of the overnight and evening periods. This asymmetry is consistent with the hydro plant's scheduled output providing substantial room to reduce generation for downward reserve, whereas upward reserve requires available upward headroom and is concentrated in periods in which reserving that flexibility is economically attractive. As the strategy becomes more conservative, the \ac{ess} is increasingly recruited for downward reserve, whereas the downward-reserve contributions of the wind and solar \ac{pv} plants decrease. This reduction follows the substantial decline in their scheduled energy production under more conservative strategies, which leaves less scheduled renewable output available to be curtailed for downward-reserve provision. The resulting shift toward storage reflects a redistribution of portfolio flexibility as the uncertainty-handling strategy becomes more conservative.

%\clearpage
\subsection{Case 2: In-Sample Economic Comparison}
\label{subsec:Case2}

This case study benchmarks the proposed \ac{plmbro} approach against the percentile-based \ac{mbro} benchmark, in which the deviation bounds are constructed from predefined percentile levels and the uncertainty budgets are manually selected by the user rather than derived from historical data. Both models are solved for the three uncertainty-handling strategies (optimistic, balanced, and pessimistic). \Cref{table:Economic_Case2} reports the economic indicators obtained directly from the robust bidding optimization, while \Cref{table:InSample_Case2} reports the in-sample assessment obtained by evaluating the fixed market bids against the historical realizations used to construct the uncertainty set. The detailed mathematical formulation used for the in-sample and out-of-sample assessments is provided in~\cite{nemati2026multi}. It should be noted that, since demand is modeled as an internal component of the \ac{rvpp} and no explicit revenue is credited for serving it, the reported profit follows the accounting convention of the objective and may take negative values; these values are therefore meaningful for comparing models and strategies rather than as absolute earnings, and a higher (less negative) value denotes a better net economic outcome.

\Cref{table:Economic_Case2} compares the two models at the bidding stage, before any uncertainty realization is observed. Expectedly, the optimized profit decreases as the operator moves toward more conservative uncertainty-handling strategies in both models because the adverse uncertainty settings reduce profitable market exposure. In the benchmark \ac{mbro} model, the nominal revenue falls from 40.76~k€{} under the optimistic strategy to 4.28 and 1.01~k€{} under the balanced and pessimistic strategies, corresponding to reductions of 89.5\% and 97.5\%, respectively. The proposed \ac{plmbro} model retains considerably larger market revenues of 20.77 and 13.38~k€{} under the same two strategies, approximately 4.9 and 13.2 times the corresponding benchmark values, while its optimized profit remains at $-0.98$ and $-5.82$~k€{}. This difference is consistent with the proposed model allocating the uncertainty budgets according to parameter-specific empirical cluster frequencies, whereas the benchmark applies the same externally specified budget to every parameter and bound within a given strategy. Under the optimistic strategy, \ac{plmbro} also attains a higher optimized profit (7.72 versus 4.08~k€{}) despite a lower nominal revenue (33.60 versus 40.76~k€{}), because its explicit market-price protection cost is 7.36~k€{} compared with 18.35~k€{} for the benchmark. The results therefore indicate a more favorable trade-off between market participation and the explicit price-protection term under the proposed calibration.

For the in-sample assessment, the optimized \ac{dam} energy bid and upward/downward \ac{srm} reserve bids are fixed, while the unit-level operating decisions are re-optimized for each historical realization using the corresponding realized market prices, renewable availability, and demand. Surplus energy is settled at the observed downward imbalance price, whereas an energy shortage is purchased at the observed upward imbalance price, with both imbalance-price series obtained from~\cite{REE2025}, following the back-testing procedure in~\cite{baringo2018day}. The realized profit is calculated as the pre-settlement profit net of the resulting imbalance cost. \Cref{table:InSample_Case2} evaluates the fixed market bids on the 92 calibration days used to construct the uncertainty set. The proposed model attains a higher average realized profit under all three strategies: 0.70 versus 0.45~k€{} under the optimistic strategy, $-0.20$ versus $-2.73$~k€{} under the balanced strategy, and $-1.68$ versus $-3.56$~k€{} under the pessimistic strategy. The corresponding improvements are 0.25, 2.53, and 1.88~k€, respectively. The source of this advantage differs across strategies. Under the optimistic strategy, \ac{plmbro} has a slightly lower pre-settlement profit than the benchmark (6.49 versus 6.66~k€{}) but reduces the average imbalance cost by 0.42~k€{}, yielding the higher realized profit. Under the balanced strategy, its imbalance cost is higher (1.13 versus 0.77~k€{}), but this is more than offset by a 2.89~k€{} improvement in pre-settlement profit (0.93 versus $-1.96$~k€{}). Under the pessimistic strategy, both components improve: the pre-settlement profit increases by 1.53~k€{} and the average imbalance cost falls from 0.64 to 0.29~k€{}. Thus, the in-sample advantage of \ac{plmbro} does not depend on systematically minimizing the imbalance cost; rather, it arises from a more favorable overall balance between the economic position established before settlement and the subsequent imbalance cost.

\begin{table}[t!]
  \centering
\caption{Case 2 -- Economic results from the robust bidding optimization for the proposed \ac{plmbro} approach and the percentile-based \ac{mbro} benchmark under the three uncertainty-handling strategies.}
  \setlength{\tabcolsep}{4pt}
  \renewcommand{\arraystretch}{1.1}
  \begin{threeparttable}
  \begin{tabular}{cclcccc}
    \toprule

    \multicolumn{1}{c}{\textbf{Model}}
     & \multicolumn{1}{c}{\textbf{Strategy}}
    && \multicolumn{1}{c}{\text{Optimized profit}}
    & \multicolumn{1}{c}{\text{Nominal revenue}}
    & \multicolumn{1}{c}{\text{Operation cost}}
    & \multicolumn{1}{c}{\shortstack{Price-protection cost}}
    \\
    \multicolumn{1}{c}{\textbf{}} & {\textbf{}} && \text{[k€]} & \text{[k€]} & \text{[k€]} & \text{[k€]}
    \\

    \specialrule{2pt}{2pt}{2pt}
    \multirow{3}{*}{MBRO} & {Optimistic}  && 4.08 & 40.76 & 18.33 & 18.35 \\ \cmidrule{2-7}
    & {Balanced}    && $-11.40$ & 4.28 & 12.59 & 3.09 \\ \cmidrule{2-7}
    & {Pessimistic} && $-12.84$ & 1.01 & 13.24 & 0.61 \\ [0.2em]

    \specialrule{1pt}{1pt}{1pt}
    \multirow{3}{*}{PL-MBRO} & {Optimistic}  && 7.72 & 33.60 & 18.52 & 7.36 \\ \cmidrule{2-7}
    & {Balanced}    && $-0.98$ & 20.77 & 12.65 & 9.10 \\ \cmidrule{2-7}
    & {Pessimistic} && $-5.82$ & 13.38 & 10.57 & 8.63 \\ [0.2em]

\bottomrule
  \end{tabular}
  \begin{tablenotes}
\footnotesize
\item The price-protection cost is the explicit market-price robustness term in~\eqref{MILP: Obj}; \ac{ndrs} and demand uncertainty affect the optimized profit through the robust operational constraints rather than through separate objective-cost terms.
\end{tablenotes}
%\vspace{4mm}
\end{threeparttable}
  \label{table:Economic_Case2}
\end{table}

\begin{table}[t!]
  \centering
    \caption{Case 2 -- In-sample economic assessment of the fixed market bids on the 92 calibration days: average pre-settlement profit, average imbalance cost, and average realized profit.}
  \setlength{\tabcolsep}{4pt}
  \renewcommand{\arraystretch}{1.1}
  \begin{threeparttable}
  \begin{tabular}{cclccc}
    \toprule

    \multicolumn{1}{c}{\textbf{Model}}
     & \multicolumn{1}{c}{\textbf{Strategy}}
    && \multicolumn{1}{c}{\shortstack{Pre-settlement profit}} 
    & \multicolumn{1}{c}{\shortstack{Imbalance cost}}
    & \multicolumn{1}{c}{\shortstack{Realized profit}}
    \\
    \multicolumn{1}{c}{\textbf{}} & {\textbf{}} && \text{[k€]} & \text{[k€]} & \text{[k€]}
    \\

    \specialrule{2pt}{2pt}{2pt}
    \multirow{3}{*}{MBRO} & {Optimistic}  && 6.66 & 6.21 & 0.45 \\ \cmidrule{2-6}
    & {Balanced}    && $-1.96$ & 0.77 & $-2.73$ \\ \cmidrule{2-6}
    & {Pessimistic} && $-2.92$ & 0.64 & $-3.56$ \\ [0.2em]

    \specialrule{1pt}{1pt}{1pt}
    \multirow{3}{*}{PL-MBRO} & {Optimistic}  && 6.49 & 5.79 & 0.70 \\ \cmidrule{2-6}
    & {Balanced}    && 0.93 & 1.13 & $-0.20$ \\ \cmidrule{2-6}
    & {Pessimistic} && $-1.39$ & 0.29 & $-1.68$ \\ [0.2em]

\bottomrule
  \end{tabular}
%\vspace{4mm}
\end{threeparttable}
  \label{table:InSample_Case2}
\end{table}

%\clearpage
\subsection{Case 3: Out-of-Sample Economic Assessment}
\label{subsec:Case3}

Case~3 evaluates the out-of-sample economic performance of the market bids produced by the two approaches. For each uncertainty-handling strategy, the \ac{dam} energy and secondary-reserve bids obtained in Case~2 are fixed and evaluated on 20 held-out historical days that were not used to construct the uncertainty set, using the same assessment and imbalance-settlement procedure described in Case~2. The held-out days belong to the same seasonal regime as the calibration data, so the assessment focuses on performance on unseen days without introducing a seasonal-regime shift. \Cref{table:OFS_Case3} reports the average pre-settlement profit, average imbalance cost, and average realized profit. This assessment complements the in-sample comparison of Case~2 by testing the economic performance of the calibrated bids on unseen historical realizations.

\Cref{table:OFS_Case3} shows that the average realized profit is negative for all model--strategy combinations on the 20 held-out days, but \ac{plmbro} attains a higher value in every comparison. 
%Its average realized profit is $-3.01$, $-5.74$, and $-7.69$~k€{} under the optimistic, balanced, and pessimistic strategies, compared with $-7.05$, $-7.85$, and $-8.66$~k€{} for the benchmark. 
The corresponding improvements for the three strategies are 4.04, 2.11, and 0.97~k€. The advantage arises primarily from the pre-settlement economic position. Under the optimistic strategy, for example, the pre-settlement profit improves from $-4.78$ to $-0.78$~k€{}, while the imbalance costs are nearly identical (2.27 and 2.23~k€{}). Under the balanced and pessimistic strategies, \ac{plmbro} incurs somewhat higher imbalance costs than the benchmark (0.38 versus 0.19~k€{} and 0.31 versus 0.16~k€{}, respectively), but the improvements in pre-settlement profit are sufficiently large to preserve a higher realized profit. Therefore, the out-of-sample advantage is driven mainly by the economic position established by the calibrated bids rather than by systematically lower imbalance costs. Together with the in-sample results, this provides evidence that the proposed calibration retains an economic advantage on unseen days within the seasonal regime considered. More generally, these results suggest that parameter-specific calibration of uncertainty bounds and budgets can be beneficial when the historical variability and occurrence patterns differ across uncertain parameters. Although this principle is applicable to other \ac{rvpp} configurations and market conditions, the magnitude of the economic improvement may depend on the resource portfolio, its flexibility, and the characteristics of the historical data.

\begin{table}[t!]
  \centering
    \caption{Case 3 -- Out-of-sample economic assessment of the fixed market bids on the 20 held-out historical days: average pre-settlement profit, average imbalance cost, and average realized profit.}
  \setlength{\tabcolsep}{6pt}
  \renewcommand{\arraystretch}{1}
  \begin{threeparttable}
  \begin{tabular}{cclccc}
    \toprule

    \multicolumn{1}{c}{\textbf{Model}}
     & \multicolumn{1}{c}{\textbf{Strategy}}
    && \multicolumn{1}{c}{\shortstack{Pre-settlement profit}}
    & \multicolumn{1}{c}{\shortstack{Imbalance cost}}
    & \multicolumn{1}{c}{\shortstack{Realized profit}}
    \\
    \multicolumn{1}{c}{\textbf{}} & {\textbf{}} && \text{[k€]} & \text{[k€]} & \text{[k€]}
    \\

    \specialrule{2pt}{2pt}{2pt}
    \multirow{3}{*}{MBRO} & {Optimistic}  && $-4.78$ & 2.27 & $-7.05$ \\ \cmidrule{2-6}
    & {Balanced}    && $-7.66$ & 0.19 & $-7.85$ \\ \cmidrule{2-6}
    & {Pessimistic} && $-8.50$ & 0.16 & $-8.66$ \\ [0.2em]

    \specialrule{1pt}{1pt}{1pt}
    \multirow{3}{*}{PL-MBRO} & {Optimistic}  && $-0.78$ & 2.23 & $-3.01$ \\ \cmidrule{2-6}
    & {Balanced}    && $-5.36$ & 0.38 & $-5.74$ \\ \cmidrule{2-6}
    & {Pessimistic} && $-7.38$ & 0.31 & $-7.69$ \\ [0.2em]

\bottomrule
  \end{tabular}
%\vspace{5mm}
\end{threeparttable}
  \label{table:OFS_Case3}
\end{table}

\section{Conclusion}
\label{sec:Conclusion}

This paper proposed \ac{plmbro}, a data-driven calibration approach for the uncertainty set of a multi-bound robust \ac{rvpp} bidding model. For each uncertain parameter---\ac{dam} and \ac{srm} prices, \ac{ndrs} production, and demand---$k$-medians clustering identifies parameter-specific probability layers. Strategy-dependent within-cluster representatives determine the corresponding deviation bounds, while the empirical probability of each cluster determines its full-horizon uncertainty budget. Once the risk-attitude quantile levels are specified, the resulting bounds and budgets are generated systematically from the historical data. Embedded in a single-level \ac{milp} for joint energy and reserve participation, the proposed calibration preserves the tractability of the underlying multi-bound robust formulation while replacing predefined percentile-based bounds and externally specified uncertainty budgets with a reproducible, parameter-specific procedure.

Three main conclusions emerge from the case studies. First, the empirical cluster probabilities reveal markedly non-uniform and parameter-specific uncertainty structures, showing that a common budget allocation across uncertain parameters does not reflect their observed frequencies. Second, on the 92 calibration days considered in the case study, \ac{plmbro} achieves a higher average realized profit than the percentile-based \ac{mbro} benchmark under all three uncertainty-handling strategies.
%, with improvements of up to 2.53~k€{}. 
These gains arise from a more favorable overall balance between the economic position established before settlement and the subsequent imbalance cost rather than from systematically minimizing imbalance costs. Third, the economic advantage persists on the 20 held-out days: \ac{plmbro} again achieves a higher average realized profit under the optimistic, balanced, and pessimistic strategies.
%, with an improvement of up to 4.04~k€{}. 
The out-of-sample advantage is driven mainly by the stronger pre-settlement economic position established by the calibrated market bids. These results provide evidence that the proposed probability-layer calibration can improve the economic performance of robust \ac{rvpp} bidding on unseen realizations within the seasonal regime considered.

Future work will extend the probability-layer framework to represent temporal dependence and cross-parameter correlations among the uncertain quantities. Further developments will also consider rolling intraday recalibration of the deviation bounds and uncertainty budgets as new observations become available.

\section*{Acknowledgments}
The authors wish to thank Comunidad de Madrid for the financial support to PREDFLEX project (TEC-2024/ECO-287), through the R\&D activity programme Tecnologías 2024.

\begin{footnotesize}
\setlength{\bibsep}{5pt}
\bibliography{refs.bib}

@article{zare2026systematic,
  author  = {Zare, Alireza and Shafie-khah, Miadreza and Siano, Pierluigi and Lazaroiu, George Cristian},
  title   = {A systematic review of virtual power plant configurations and their interaction with electricity, carbon, and flexibility markets},
  journal = {Renewable and Sustainable Energy Reviews},
  volume  = {226},
  pages   = {116448},
  year    = {2026}
}

@article{xu2026bidding,
  author  = {Xu, Mengchao and Yang, Xiyun and Bu, Dongdong and Wei, Chang},
  title   = {Two-stage distributionally robust bidding strategy of virtual power plants participating in electricity markets considering {WT-PV} complementarity and multiple uncertainties},
  journal = {Applied Energy},
  volume  = {423},
  pages   = {128438},
  year    = {2026}
}

@article{kong2026coordinated,
  author  = {Kong, Xiangyu and Chen, Yi and Zhang, Xiyuan and Liu, Gaohua and Zhang, Peirong and Zhang, Delong},
  title   = {Coordinated decision-making method for virtual power plant participating in multiple power spot markets},
  journal = {Energy},
  volume  = {362},
  pages   = {142183},
  year    = {2026}
}

@article{bian2026cvar,
  author  = {Bian, Jing and Cheng, Jianing and Wang, He and Li, Guoqing and Li, Shiqiang},
  title   = {{CVaR}-based bi-level bidding strategy for {VPP} to participate in energy-regulation-carbon markets},
  journal = {International Journal of Electrical Power \& Energy Systems},
  volume  = {174},
  pages   = {111555},
  year    = {2026}
}

@article{pleite2026bidding,
  author  = {Pleite-Paniagua, Andr{\'e}s and Garc{\'i}a-Cerezo, {\'A}lvaro and Carri{\'o}n, Miguel},
  title   = {Building day-ahead market bidding and offering curves of a virtual power plant participating in a manual frequency restoration reserve-based demand response program},
  journal = {Sustainable Energy, Grids and Networks},
  volume  = {47},
  pages   = {102426},
  year    = {2026}
}

@article{chang2025game,
  author  = {Chang, Weiguang and Yang, Qiang},
  title   = {Game-theoretic risk-averse day-ahead optimal bidding strategy of virtual power plant aggregated with heterogeneous distributed resources},
  journal = {Energy},
  volume  = {336},
  pages   = {138087},
  year    = {2025}
}

@article{bo2026adaptive,
  author  = {Bo, Yaolong and Yuan, Mengtong and Zhu, Chengzhi and Xia, Yanghong and Wei, Wei},
  title   = {Adaptive look-ahead economic dispatch with reinforcement learning for virtual power plant considering uncertainty},
  journal = {Applied Energy},
  volume  = {424},
  pages   = {128486},
  year    = {2026}
}

@article{wang2026drmpc,
  author  = {Wang, Xiaobin and Li, Qi and Zhai, Junyi and Jiang, Yuning and Wang, Sheng and Wang, Jianxiao},
  title   = {Flexible operation of virtual power plant enabled integrated electricity-heating system under multiple uncertainties via distributionally robust model predictive control},
  journal = {Applied Energy},
  volume  = {404},
  pages   = {127177},
  year    = {2026}
}

@article{gela2026metamodel,
  author  = {Gela, Temesgen Teshager and Mohandes, Mohamed and Al-Shaikhi, Ali and Al-Ismail, Fahad Saleh and Gebril, Abdelaziz M. and Zalhaf, Amr S.},
  title   = {Metamodel-assisted stochastic {Stackelberg} coordination of multiple virtual power plants under joint wind, load, and price uncertainty},
  journal = {Sustainable Energy, Grids and Networks},
  volume  = {47},
  pages   = {102491},
  year    = {2026}
}

@article{lv2026drl,
  author  = {Lv, You and Liao, Conglin and Tian, Helu and Zhao, Chen and Wang, Xuedan and Liu, Pin and Fang, Fang},
  title   = {Multi-objective optimal scheduling of virtual power plants using deep reinforcement learning to address uncertainty},
  journal = {International Journal of Electrical Power \& Energy Systems},
  volume  = {180},
  pages   = {112059},
  year    = {2026}
}

@article{nemati2026multi,
title = {A multi-bound robust optimization approach for renewable-based {VPP} market participation considering intra-hourly uncertainty exposure},
journal = {Applied Energy},
volume = {426},
pages = {128765},
year = {2026},
issn = {0306-2619},
author={Nemati, Hadi and Ortega, {\'A}lvaro and Lobato, Enrique and Rouco, Luis},
}

@article{li2026risk,
  title={Risk-averse co-optimization of virtual power plants in multi-markets considering scheduling costs and uncertainties of controllable loads},
  author={Li, Qiang and Dong, Fuxiang and Jiang, Bo and Liu, Jinfu and Yan, Peigang and Yu, Daren},
  journal={Applied Energy},
  volume={409},
  pages={127496},
  year={2026},
  publisher={Elsevier}
}

@misc{EC_Energy_Market,
  author       = {{European Commission}},
  title        = {{Electricity Market Design}},
  howpublished = {\url{https://energy.ec.europa.eu/topics/markets-and-consumers/electricity-market-design_en}},
  year         = {2025}
}

@misc{omie_homepage,
  author       = {{Operador del Mercado Ibérico de Energía (OMIE)}},
  howpublished = {\url{https://www.omie.es/en}},
  year         = {2025}
}

@misc{REE2025,
  author       = {{Red Eléctrica de España (REE)}},
  howpublished = {https://www.esios.ree.es/},
  year         = {2026}
}

@article{mei2024two,
  title={A two-step optimization model for virtual power plant participating in spot market based on energy storage power distribution considering comprehensive forecasting error of renewable energy output},
  author={Mei, Shufan and Tan, Qinliang and Trivedi, Anupam and Srinivasan, Dipti},
  journal={Applied Energy},
  volume={376},
  pages={124234},
  year={2024},
  publisher={Elsevier}
}

@article{wang2024two,
  title={Two-stage distributionally robust offering and pricing strategy for a price-maker virtual power plant},
  author={Wang, Jun and Xu, Jian and Wang, Jingjing and Ke, Deping and Yao, Liangzhong and Zhou, Yue and Liao, Siyang},
  journal={Applied Energy},
  volume={363},
  pages={123005},
  year={2024},
  publisher={Elsevier}
}

@article{yuanyuan2023distributionally,
  title={Distributionally robust comprehensive declaration strategy of virtual power plant participating in the power market considering flexible ramping product and uncertainties},
  author={Yuanyuan, Zhang and Huiru, Zhao and Bingkang, Li},
  journal={Applied Energy},
  volume={343},
  pages={121133},
  year={2023},
  publisher={Elsevier}
}

@article{li2025internal,
  title={Internal and external coordinated distributionally robust bidding strategy of virtual power plant operator participating in day-ahead electricity spot and peaking ancillary services markets},
  author={Li, Wanying and Dong, Fugui and Ji, Zhengsen and Wang, Peijun},
  journal={Applied Energy},
  volume={386},
  pages={125514},
  year={2025},
  publisher={Elsevier}
}

@article{afzali2025flexibility,
  title={A flexibility-oriented bidding strategy for virtual power plants incorporating local energy communities: A bi-level stochastic-robust methodology},
  author={Afzali, Soheil and Moghaddam, Mohsen Parsa and Sheikh-El-Eslami, Mohammad Kazem and Zamani, Reza},
  journal={Applied Energy},
  volume={399},
  pages={126355},
  year={2025},
  publisher={Elsevier}
}

@article{ma2025data,
  title={Data-Driven interval robust optimization method of {VPP} Bidding strategy in spot market under multiple uncertainties},
  author={Ma, Ying and Li, Zhen and Liu, Ruyi and Liu, Bin and Yu, Samson S and Liao, Xiaozhong and Shi, Peng},
  journal={Applied Energy},
  volume={384},
  pages={125366},
  year={2025},
  publisher={Elsevier}
}

@article{zhu2025low,
  title={Low-carbon and robust economic scheduling of virtual power plants considering multiple uncertainties},
  author={Zhu, Yongcan and Wei, Naying and Kang, Junjun and Tian, Yi},
  journal={Sustainable Energy, Grids and Networks},
  pages={102104},
  year={2025},
  publisher={Elsevier}
}

@article{nokandi2023three,
  title={A three-stage bi-level model for joint energy and reserve scheduling of {VPP} considering local intraday demand response exchange market},
  author={Nokandi, Ehsan and Vahedipour-Dahraie, Mostafa and Goldani, Saeed Reza and Siano, Pierluigi},
  journal={Sustainable Energy, Grids and Networks},
  volume={33},
  pages={100964},
  year={2023},
  publisher={Elsevier}
}

@article{shang2025uncertainty,
  title={Uncertainty-output virtual power plant participates in multi-electricity market considering the improved Shapley value distribution method},
  author={Shang, Yi and Li, Xiaolan and Xu, Tianqi and Cui, Lin},
  journal={International Journal of Electrical Power \& Energy Systems},
  volume={165},
  pages={110462},
  year={2025},
  publisher={Elsevier}
}

@article{li2023robust,
  title={Robust bidding strategy for multi-energy virtual power plant in peak-regulation ancillary service market considering uncertainties},
  author={Li, Yong and Deng, Youyue and Wang, Yahui and Jiang, Lin and Shahidehpour, Mohammad},
  journal={International Journal of Electrical Power \& Energy Systems},
  volume={151},
  pages={109101},
  year={2023},
  publisher={Elsevier}
}

@article{wu2024two,
  title={Two-stage distributionally robust optimal operation of rural virtual power plants considering multi correlated uncertainties},
  author={Wu, Shenglei and Wang, Yong and Liu, Lurao and Yang, Zheng and Cao, Qiang and He, Huijun and Cao, Yiyao},
  journal={International Journal of Electrical Power \& Energy Systems},
  volume={161},
  pages={110173},
  year={2024},
  publisher={Elsevier}
}

@article{du2025optimal,
  title={Optimal bidding strategy for virtual power plant participating in joint energy and reserve market considering physical-economic feasible operation region},
  author={Du, Haoming and Hu, Junjie and Fang, Yuxuan and Xie, Dongliang},
  journal={Energy},
  pages={138410},
  year={2025},
  publisher={Elsevier}
}

@article{liu2025two,
  title={Two-stage robust optimization of a virtual power plant considering a refined demand response},
  author={Liu, Jinpeng and Peng, Jinchun and Liu, Hushihan and Deng, Jiaming and Song, Xiaohua},
  journal={Energy},
  volume={322},
  pages={135560},
  year={2025},
  publisher={Elsevier}
}

@article{feng2024optimal,
  title={Optimal energy scheduling of virtual power plant integrating electric vehicles and energy storage systems under uncertainty},
  author={Feng, Jie and Ran, Lun and Wang, Zhiyuan and Zhang, Mengling},
  journal={Energy},
  volume={309},
  pages={132988},
  year={2024},
  publisher={Elsevier}
}

@article{wang2023optimal,
  title={Optimal self-scheduling for a multi-energy virtual power plant providing energy and reserve services under a holistic market framework},
  author={Wang, Jian and Ilea, Valentin and Bovo, Cristian and Xie, Ning and Wang, Yong},
  journal={Energy},
  volume={278},
  pages={127903},
  year={2023},
  publisher={Elsevier}
}

@article{siqin2025two,
  title={Two-stage distributionally robust bidding strategies for multi-energy virtual power plants in energy, reserve and carbon markets},
  author={Siqin, Zhuoya and Diao, Ruisheng and Zhao, Mingrui},
  journal={Energy},
  pages={137455},
  year={2025},
  publisher={Elsevier}
}

@article{esfahani2024stochastic,
  title={A stochastic-robust aggregation strategy for {VPP} to participate in the frequency regulation market via backup batteries},
  author={Esfahani, Moein and Alizadeh, Ali and Cao, Bo and Kamwa, Innocent and Xu, Minghui},
  journal={Journal of Energy Storage},
  volume={98},
  pages={113057},
  year={2024},
  publisher={Elsevier}
}

@article{feng2025optimal,
  title={Optimal bidding strategy for virtual power plant in multiple markets considering integrated demand response and energy storage},
  author={Feng, Jie and Ran, Lun and Wang, Zhiyuan and Zhang, Mengling},
  journal={Journal of Energy Storage},
  volume={124},
  pages={116706},
  year={2025},
  publisher={Elsevier}
}

@article{wang2024optimal,
  title={Optimal scheduling strategy for virtual power plants with aggregated user-side distributed energy storage and photovoltaics based on {CVaR}-distributionally robust optimization},
  author={Wang, Yushen and Huang, Weiliang and Chen, Haoyong and Yu, Zhiwen and Hu, Linlin and Huang, Yuxiang},
  journal={Journal of Energy Storage},
  volume={86},
  pages={110770},
  year={2024},
  publisher={Elsevier}
}

@article{falabretti2023scheduling,
  title={Scheduling and operation of {RES}-based virtual power plants with e-mobility: A novel integrated stochastic model},
  author={Falabretti, Davide and Gulotta, Francesco and Siface, Dario},
  journal={Int. J. Electr. Power Energy Syst.},
  volume={144},
  pages={108604},
  year={2023},
  publisher={Elsevier}
}

@article{yang2023optimal,
  title={Optimal bidding strategy of renewable-based virtual power plant in the day-ahead market},
  author={Yang, Chen and Du, Xiao and Xu, Dan and Tang, Junjie and Lin, Xingyu and Xie, Kaigui and Li, Wenyuan},
  journal={Int. J. Electr. Power Energy Syst.},
  volume={144},
  pages={108557},
  year={2023},
  publisher={Elsevier}
}

@article{shafiekhani2022optimal,
  title={Optimal bidding strategy of a renewable-based virtual power plant including wind and solar units and dispatchable loads},
  author={Shafiekhani, Morteza and Ahmadi, Abdollah and Homaee, Omid and Shafie-khah, Miadreza and Catal{\~a}o, Jo{\~a}o PS},
  journal={Energy},
  volume={239},
  pages={122379},
  year={2022},
  publisher={Elsevier}
}

@article{bertsimas04,
  title={The price of robustness},
  author={Bertsimas, D. and Sim, M.},
  journal={Operations Research},
  volume={52},
  number={1},
  pages={35--53},
  year={2004},
}

@article{baringo2018day,
  title={Day-ahead self-scheduling of a virtual power plant in energy and reserve electricity markets under uncertainty},
  author={Baringo, Ana and Baringo, Luis and Arroyo, Jose M},
  journal={IEEE Transactions on Power Systems},
  volume={34},
  number={3},
  pages={1881--1894},
  year={2018},
  publisher={IEEE}}

@inproceedings{busing2012new,
  title={New results about multi-band uncertainty in robust optimization},
  author={B{\"u}sing, Christina and D’andreagiovanni, Fabio},
  booktitle={International symposium on experimental algorithms},
  pages={63--74},
  year={2012},
  organization={Springer}
}

@article{morales2018robust,
  title={Robust unit commitment with dispatchable wind power},
  author={Morales-Espa{\~n}a, Germ{\'a}n and Lorca, {\'A}lvaro and de Weerdt, Mathijs M},
  journal={Electric Power Systems Research},
  volume={155},
  pages={58--66},
  year={2018},
  publisher={Elsevier}
}

@article{zhang2022frequency,
  title={A frequency security constrained scheduling approach considering wind farm providing frequency support and reserve},
  author={Zhang, Zhi and Zhou, Ming and Wu, Zhaoyuan and Liu, Siwei and Guo, Zun and Li, Gengyin},
  journal={IEEE Transactions on Sustainable Energy},
  volume={13},
  number={2},
  pages={1086--1100},
  year={2022},
  publisher={IEEE}
}

@article{yin2021state,
  title={State-of-the-art short-term electricity market operation with solar generation: A review},
  author={Yin, Shengfei and Wang, Jianhui and Li, Zhengshuo and Fang, Xin},
  journal={Renewable and Sustainable Energy Reviews},
  volume={138},
  pages={110647},
  year={2021},
  publisher={Elsevier}
}

@article{carrion2006computationally,
  title={A computationally efficient mixed-integer linear formulation for the thermal unit commitment problem},
  author={Carri{\'o}n, Miguel and Arroyo, Jos{\'e} M},
  journal={IEEE Transactions on power systems},
  volume={21},
  number={3},
  pages={1371--1378},
  year={2006},
  publisher={IEEE}
}

@article{nemati2025segan,
  title={Flexible Robust Optimal Bidding of Renewable Virtual Power Plants in Sequential Markets under Asymmetric Uncertainties},
  author={Nemati, Hadi and S{\'a}nchez-Mart{\'\i}n, Pedro and Ortega, {\'A}lvaro and Sigrist, Lukas and Lobato, Enrique and Rouco, Luis},
  journal={Sustainable Energy, Grids and Networks},
  pages={101801},
  year={2025},
  publisher={Elsevier}
}
\bibliographystyle{elsarticle-num.bst} 

\end{footnotesize}

\end{document}